\documentclass[useAMS,usenatbib]{mn2e}
\usepackage{times}
\usepackage{graphics,epsfig}
\usepackage{graphicx}
\usepackage{amssymb}
\usepackage{txfonts}

\newcommand{\kms}{km\,s$^{-1}$} 
\newcommand{\Ts}   {\ensuremath{{\rm T}_{\rm s}}}
\newcommand{\Td}   {\ensuremath{{\rm T}_{\rm D}}}
\newcommand{\Tk}   {\ensuremath{{\rm T}_{\rm k}}}

\newcommand{\dnhi} {\ensuremath{\Delta{\rm N(HI)}}}

\newcommand{\cm}{cm$^{-2}$}
\newcommand{\hi}{H\,{\sc i}}
\newcommand{\hii}{H\,{\sc i}-21\,cm}

\title[Temperature of the ISM]{The temperature of the diffuse H\,{\sc i} in the Milky Way II: Gaussian decomposition of the H\,{\sc i} 21cm absorption spectra}
\author[N. Roy et al.]{Nirupam Roy$^{1}$\thanks{E-mail: nirupam@mpifr-bonn.mpg.de~(NR);~~~ nkanekar@ncra.tifr.res.in~(NK); chengalu@ncra.tifr.res.in~(JNC)}, 
Nissim Kanekar$^{2}$\footnotemark[1] and Jayaram N. Chengalur $^{2}$\footnotemark[1]\\
$^{1}$Max-Planck-Institut f\"{u}r Radioastronomie, Auf dem H\"{u}gel 69, D-53121, Bonn, Germany\\
$^{2}$National Centre for Radio Astrophysics, TIFR, Post Bag 3, Ganeshkhind, Pune 411 007, India}

\begin{document}
\date{Accepted yyyy month dd. Received yyyy month dd; in original form yyyy month dd}

\pagerange{\pageref{firstpage}--\pageref{lastpage}} 
\pubyear{2013}

\maketitle

\label{firstpage}

\begin{abstract}
We discuss physical conditions in Galactic neutral hydrogen based on deep, 
high velocity resolution interferometric H\,{\sc i} 21cm absorption spectroscopy towards 
33 compact extra-galactic radio sources. The H\,{\sc i} 21cm optical depth spectra have 
root-mean-square noise values $\lesssim 10^{-3}$ per 1~km~s$^{-1}$ velocity 
channel, i.e., sufficiently sensitive to detect H\,{\sc i} 21cm absorption by 
the warm neutral medium (WNM). Comparing these spectra with H{\sc i} 21cm emission 
spectra from the Leiden-Argentine-Bonn (LAB) survey, we show that some of the absorption 
detected on most sightlines must arise in gas with temperatures higher than that in the 
stable cold neutral medium (CNM). A multi-Gaussian decomposition of 30 of the H\,{\sc i} 
21cm absorption spectra yielded very few components with line widths in the temperature range 
of stable WNM, with no such WNM components detected for sixteen of the thirty sightlines.
We find that some of the detected H\,{\sc i} 21cm absorption along thirteen of these 
sightlines must arise in gas with spin temperatures larger than the CNM range. For these 
sightlines, we use very conservative estimates of the CNM spin temperature and the 
non-thermal broadening to derive strict upper limits to the gas column densities
in the CNM and WNM phases. Comparing these upper limits to the total H\,{\sc i} 
column density, we find that typically at least 28\% of the gas must have temperatures 
in the thermally unstable range ($200-5000$~K). Our observations hence robustly indicate 
that a significant fraction of the gas in the Galactic interstellar medium has 
temperatures outside the ranges expected for thermally stable gas in two-phase models.
\end{abstract}


\section{Introduction}
\label{sec:intro}

In the thermal steady-state model for neutral hydrogen (\hi) in the Galactic 
interstellar medium (ISM), two distinct stable phases coexist in pressure 
equilibrium \citep[e.g.][]{field65,field69,wolfire95,wolfire03}. These are 
(1)~a cold phase (the cold neutral medium, CNM), with high density ($n \approx 
10 - 100$ cm$^{-3}$) and a high \hii\ optical depth that gives rise to the 
narrow absorption features seen towards continuum radio sources, and (2)~a 
warm diffuse phase (the warm neutral medium, WNM) with low density ($n \approx 
0.1 - 1$ cm$^{-3}$), which contributes to the \hi\ emission, but is extremely 
difficult to detect in absorption due to its low optical depth. The models 
\citep[e.g.][]{wolfire95} find that gas in the stable CNM phase would have 
kinetic temperatures $T_k \sim 40-200$ K, while gas in the stable WNM phase 
would have kinetic temperatures of $T_k \sim 5000 - 8000$\,K. Over the last 
few decades, a number of \hii\ studies have established that the ISM indeed 
contains cold atomic gas in the temperature range $\sim 40 - 200$\,K, 
consistent with theoretical expectations 
\citep[e.g.][]{clark62,rad72,dickey78,heiles03a,roy06}. However, observational 
estimates of WNM temperature are quite rare \citep{heiles03b,kanekar03b} and 
very little is as yet known about physical conditions in the WNM.

In the two-phase models, \hi\ at intermediate temperatures is unstable and 
such gas is expected to quickly evolve into one of the stable phases. Gas at 
intermediate temperatures is thus expected to exist ``only as a transient 
phenomenon'' \citep{field69}. The thermal timescale derived from the heating 
or cooling rates is $\sim 2\times10^4$~yr for the CNM, and about $100$ times 
larger for the WNM \citep{wolfire95}. On the other hand, the timescale between 
physical disturbances in the ISM (i.e. the typical time interval between 
significant pressure fluctuations) due to the propagation of supernova shocks 
is $\approx 4\times10^5$~yr \citep{mckee77,wolfire95}. Thus, while the CNM can 
be assumed to be in the thermal steady state, equilibrium conditions may not 
prevail in the WNM. However, \citet{wolfire95} argue that physical conditions 
in the WNM will evolve towards the steady state, and the steady state 
conditions could be thought to ``represent an average of the conditions to be 
expected in the actual WNM''.

Recently, there have been suggestions, based both on \hii\ observations 
and simulations, that a significant fraction of the Galactic \hi\ may have kinetic 
temperatures in the unstable range, $500 - 5000$ K \citep{heiles03a,heiles03b,kanekar03b}. 
Numerical simulations of the ISM suggest that dynamical processes like turbulence may 
drive \hi\ from the stable CNM or WNM phases to the thermally unstable phase 
\citep[e.g.][]{audit05,saury13}. This could result in significant amounts of 
unstable neutral gas, unlike the case of the standard two-phase models. 
Observational evidence for the presence of unstable gas stems from the work of 
\citet{heiles03a}, who carried out high velocity resolution \hii\ absorption and 
emission studies towards 79 compact radio sources with the Arecibo telescope. 
They then modelled the \hii\ absorption and emission spectra as the sum of 
thermally-broadened Gaussian components to estimate the kinetic temperature and 
fraction of gas in the different temperature phases. The unstable phase was found 
to make up nearly 50\% of the \hi\ along their sightlines \citep{heiles03b}. 
However, the single-dish spectra used in their analysis are subject to large systematic 
effects, both due to contamination from stray radiation coming through the large 
sidelobes of the Arecibo telescope and the fact that single-dish \hii\ absorption 
spectra must be corrected for the \hi\ emission in the beam. The sidelobe contamination 
in the emission spectra is especially important here, because the results pertaining 
to gas in the unstable phase were based on the fits to the emission spectra. 

Conversely, \citet{kanekar03b} carried out high velocity resolution {\it 
interferometric} \hii\ absorption studies along two sightlines using the 
Australia Telescope Compact Array (ATCA), with multi-Gaussian fits to the 
absorption spectra alone [see also \citet{lane00} and \citet{kanekar01b} for a 
similar approach in damped Lyman-$\alpha$ systems]. \citet{kanekar03b} found 
that a significant fraction of gas is in the thermally unstable phase; however 
the sample was small, with only two lines of sight. Such deep interferometric 
observations allow the possibility of obtaining an ``clean'' determination of 
the \hii\ absorption profile, by resolving out the \hii\ emission, and thus, 
of directly detecting the WNM in absorption. We have recently begun a large 
observational program to probe neutral atomic gas in the ISM through 
interferometric \hii\ absorption studies with the Westerbork Synthesis Radio 
Telescope (WSRT) and the Giant Metrewave Radio Telescope (GMRT), following the 
approach of \citet{kanekar03b}. The observations, data analysis, spectra and 
integrated properties for each of the 34 observed sightlines were presented by 
\citet[][hereafter, Paper I]{roy13a}. In this paper, we describe the 
multi-Gaussian parametrisation of the 33 absorption spectra with detected 
\hii\ absorption, including the two ATCA targets of \citet{kanekar03b}, and 
their implications for physical conditions in the interstellar medium. 

\section{Spin, kinetic and doppler Temperatures in the ISM}
\label{sec:tk-ts}

In Paper I, we discussed in detail the spin temperature $\Ts$ and the kinetic 
temperature $\Tk$, the two main temperatures used to characterize the neutral 
atomic phases of the interstellar medium. We will not repeat this discussion 
here, but will simply assume that $\Ts \approx \Tk$ for the CNM and that $\Ts 
\lesssim \Tk$ in the other temperature phases, with $\Ts \leq 5000$~K 
\citep[e.g.][]{liszt01}. The spin temperature along a sightline is estimated 
from the ratio of the total \hi\ column density to the velocity-integrated 
\hii\ optical depth. The \hi\ column density can be estimated either from the 
\hii\ emission spectrum along a neighbouring sightline 
\citep[e.g.][]{heiles03a} or from an absorption spectrum in the damped 
Lyman-$\alpha$ line \citep[e.g.][]{wakker11}, while the \hii\ optical depth 
is, of course, measured from \hii\ absorption studies towards a background 
radio source. For an arbitrary sightline, the spin temperature obtained from 
the above procedure is the column density weighted harmonic mean of the spin 
temperatures of individual ``clouds'' along the sightline, and does not 
provide direct information on the gas temperature. For large \hii\ optical 
depths, there are further complications because one does not {\it a priori} 
know the distribution of clouds along the sightline, making it difficult to 
correct for self-absorption in the \hii\ emission spectrum while estimating 
the total \hi\ column density . Attempts to fit for this distribution along 
the line of sight have so far not been successful \citep[e.g.][]{heiles03a}. A 
variation of the above procedure is to decompose the absorption and emission 
spectra into Gaussian components and then determine the spin temperature of 
individual components \citep[e.g.][]{dickey78,payne82,heiles03a}. This has so 
far only been successful for the cold phase, due to the high spectral dynamic 
range required to detect the WNM in absorption. Even here, there are 
complications because one is measuring the emission profile along a different 
line of sight (and generally a larger field of view) than that for the 
absorption profile, and small scale variations in the \hi\ distribution could 
invalidate the analysis.

The kinetic temperature \Tk\ is the temperature that characterizes the 
velocity distribution of the \hi\ atoms on small scales (i.e. scales on which 
bulk motions are negligible). Because these scales are generally inaccessible 
to observations, the gas kinetic temperature is instead typically estimated 
from the ``doppler temperature'' ($\Td$), which is derived from the 
full-width-at-half-maximum (FWHM) of an \hii\ component by the relation $\Td = 
21.855 \times \Delta V^2$\,K, where $\Delta V$  is the FWHM in \kms. If the 
line width is dominated by thermal broadening, with negligible contributions 
from turbulence, bulk motions, etc, one would have $\Tk = \Td$. However, in 
general, 
\begin{equation}
  \Td = \Tk + {m_H v_{turb}^2 \over k_B} \:\:,
\label{eqn:vturb}
\end{equation}
where $v_{turb}$ gives the contribution to the line width from non-thermal 
motions, $m_H$ is the mass of the hydrogen atom and $k_B$ is the Boltzmann 
constant. If such non-thermal broadening is negligible, one can use \Td\ and 
\Tk\ interchangeably. In the general case, the doppler temperature only 
provides an upper limit to the kinetic temperature.

To summarize, in \hii\ absorption spectra, both the spin and doppler 
temperatures affect the line profile; the spin temperature determines the 
total integrated optical depth, while the doppler temperature (i.e. the 
combination of the kinetic temperature and any non-thermal motions) determines 
the line width. If one assumes that non-thermal contributions to the line 
widths are small \citep[e.g.][]{heiles03a}, one can estimate the kinetic 
temperature from the line profile.  

The main differences between the present work and the analysis of 
\citet{heiles03a} is that (i)~we obtain our \hii\ absorption spectra from high 
angular resolution interferometric observations, unaffected by the \hii\ 
emission, (ii)~we carry out the multi-Gaussian decomposition for the \hii\ 
absorption profiles only, and (iii)~we use these decompositions to identify 
lines of sight along which there appears to be very little of the classical 
WNM, but the conclusions that we draw about the fractions of CNM and WNM are 
independent of the assumption that the doppler temperature approximates the 
kinetic temperature.

\section{The sample}
\label{sec:sample}

We have used the Very Large Array (VLA) calibrator list\footnote{A description 
of the VLA calibrator list can be obtained at 
https://science.nrao.edu/facilities/vla/docs/manuals/cal .} to select a sample 
of bright, compact extra-galactic background radio sources for use as targets 
for deep \hii\ absorption spectroscopy to probe physical conditions in the 
intervening Galactic neutral hydrogen. 32 such compact sources (all with 
1.4~GHz flux densities $> 3.5$~Jy) were observed with the Westerbork Synthesis 
Radio Telescope (WSRT) and the Giant Metrewave Radio Telescope (GMRT) between 
2005 and 2008; the observations, data analysis and spectra are described in 
detail in Paper I. The velocity resolutions of the spectra were $\approx 
0.41$~\kms\ (GMRT) and $\approx 0.26$~\kms\ (WSRT), except for two WSRT 
targets, B1328+254 and B1328+307, which had a resolution of $\approx 
0.52$~\kms. The observations were designed to detect the warm neutral medium 
in absorption, even for low \hi\ column densities, $\ge 2 \times 10^{20}$~\cm, 
and high spin temperatures $\Ts \approx 5000$~K. The root-mean-square (RMS) 
optical depth noise on the spectra was $\lesssim 10^{-3}$ at a resolution of 
1~\kms. \hii\ absorption was detected along 31 sightlines, with only a single 
non-detection, towards B0438$-$436 (Paper~I).

In addition, two similar bright, compact sources were observed by 
\citet{kanekar03b} with the ATCA, with a velocity resolution of $\approx 
0.41$~\kms\ and an RMS optical depth noise of $\approx 10^{-3}$ per 
$0.41$~\kms. We include these two sources with the 31 sources of Paper~I with 
detected \hii\ absorption in the following analysis. The full sample 
considered in this paper thus consists of 33~compact sources with high 
velocity resolution ($\approx 0.26 - 0.52$~\kms) interferometric \hii\ 
absorption spectra of RMS optical depth $\lesssim 10^{-3}$ per 1~\kms\ 
velocity channel. We have also obtained \hii\ emission spectra along 
neighbouring sightlines from the Leiden-Argentine-Bonn survey 
\citep{kalberla05,bajaja05}, at a velocity resolution of $\approx 1.03$~\kms. 

\begin{table}
\begin{center}
\caption{The full sample of 34 sources}
\label{table:sample}
 \begin{tabular}{lcccc}
\hline
Source      &  N(\hi) &  $\int \tau {\rm dV}$ & $\Delta V_{90}$ \\  
            &  $\times 10^{20}$\,\cm\ & \kms\  & \kms\ \\
            &  & & \\
\hline
  B0023-263 & $  1.623 \pm 0.013 $ &   $0.0254 \pm 0.0050$ &$101$ \\   
  B0114-211 & $  1.421 \pm 0.019 $ &   $0.1354 \pm 0.0054$ &$~61$ \\  
  B0117-155 & $  1.452 \pm 0.017 $ &   $0.0307 \pm 0.0042$ &$~36$ \\  
  B0134+329 & $  4.330 \pm 0.019 $ &   $~0.443 \pm 0.002$  &$~43$ \\  
  B0202+149 & $  4.803 \pm 0.016 $ &   $0.7472 \pm 0.0047$ &$~24$ \\  
  B0237-233 & $  2.132 \pm 0.020 $ &   $0.2938 \pm 0.0040$ &$~32$ \\  
  B0316+162 &  $ 10.09 \pm 0.62  $ &   $~2.964 \pm 0.004$  &$~24$ \\  
  B0316+413 & $ 13.64  \pm 0.51  $ &   $~1.941 \pm 0.003$  &$~49$ \\  
  B0355+508 & $ 115    \pm 43    $ &   $45.8 \pm 1.1$      &$~59$ \\
  B0404+768 & $ 11.06  \pm 0.42  $ &   $~1.945 \pm 0.005$  &$124$ \\  
  B0407-658 & $  3.408 \pm 0.030$  &   $0.5481 \pm 0.0070$ &$~64$ \\  
  B0429+415 & $ 41.3  \pm 1.6    $ &   $10.879 \pm 0.007$  &$~66$ \\
  B0438-436 &  $ 1.319 \pm 0.012 $ &   $~~~~~< 0.02~~~~~$  &$~63$ \\
  B0518+165 & $ 23.6  \pm 5.4    $ &   $~6.241 \pm 0.007$  &$~43$ \\  
  B0531+194 &  $ 28.8  \pm 1.5   $ &   $~4.062 \pm 0.005$  &$~32$ \\  
  B0538+498 & $ 21.31  \pm 0.94  $ &   $~5.618 \pm 0.003$  &$~71$ \\  
  B0831+557 & $ 4.499  \pm 0.020 $ &   $~0.483 \pm 0.005$  &$~75$ \\  
  B0834-196 &  $ 6.86  \pm 0.21  $ &   $~0.973 \pm 0.005$  &$~68$ \\  
  B0906+430 & $  1.283 \pm 0.020 $ &   $0.0512 \pm 0.0030$ &$~99$ \\  
  B1151-348 &  $ 7.06  \pm 0.12  $ &   $~0.714 \pm 0.004$  &$125$ \\  
  B1245-197 &  $ 3.711 \pm 0.017 $ &   $~0.158 \pm 0.005$  &$~55$ \\  
  B1328+254 & $  1.108 \pm 0.019 $ &   $0.0214 \pm 0.0031$ &$~67$ \\  
  B1328+307 & $  1.221 \pm 0.016 $ &   $0.0717 \pm 0.0018$ &$~54$ \\  
  B1345+125 &  $ 1.920 \pm 0.011 $ &   $~0.305 \pm 0.005$  &$~34$ \\  
  B1611+343 & $  1.360 \pm 0.019 $ &   $0.0187 \pm 0.0027$ &$~66$ \\  
  B1641+399 & $  1.078 \pm 0.018 $ &   $0.0088 \pm 0.0024$ &$~75$ \\  
  B1814-637 & $  6.66 \pm 0.27 $   &   $0.9974 \pm 0.0067$ &$~39$ \\  
  B1827-360 &  $ 8.13  \pm 0.40  $ &   $~1.542 \pm 0.003$  &$~58$ \\  
  B1921-293 &  $ 7.44  \pm 0.41  $ &   $~1.446 \pm 0.006$  &$~53$ \\  
  B2050+364 & $ 28.53  \pm 0.96  $ &   $~3.024 \pm 0.010$  &$~87$ \\  
  B2200+420 & $ 18.59  \pm 0.98  $ &   $~3.567 \pm 0.016$  &$~91$ \\  
  B2203-188 & $ 2.381  \pm 0.011 $ &   $0.2483 \pm 0.0042$ &$~48$ \\  
  B2223-052 &  $ 4.61  \pm 0.22  $ &   $~1.034 \pm 0.003$  &$~23$ \\  
  B2348+643 & $ 88     \pm 16    $ &   $32.517 \pm 0.031$  &$103$ \\
\hline
\end{tabular}
\end{center}
\begin{flushleft}
\end{flushleft}
\end{table}

The integrated properties along all sightlines of the sample are presented in 
Paper~I \citep[see also][]{kanekar11a}. For completeness, a few of these properties are summarized here in 
Table~\ref{table:sample}: the columns of this table contain (1)~the source 
name (using the B1950 convention), (2)~the total \hi\ column density along the 
sightline, N(\hi), using the isothermal estimate \citep{chengalur13}, (3)~the 
integrated \hii\ optical depth, $\int \tau {\rm dV}$, in \kms, and (4)~the 
velocity extent containing 90\% of the \hii\ emission, $\Delta V_{\rm 90}$, in 
\kms. 

As shown in \citet{chengalur13}, the isothermal estimate, in general, provides 
a significantly more accurate estimate of the total H~{\sc i} column density 
than the usually used optically-thin estimate. We also emphasize that (despite 
the name) the isothermal estimate does not assume that all of the gas in the 
Galaxy is at a single temperature, but rather that it uses the total optical 
depth and brightness temperature measured along the line of sight to find the 
temperature that leads to the best estimate of the total H~{\sc i} column 
density. The simulations in \citet{chengalur13} show that this estimate works 
well even at very large optical depths as well as in situations where the 
temperature varies significantly along the line of sight. We note also that 
for the lines of sight that we consider here, the difference between the 
isothermal estimate and the optically thin estimate is $< 30\%$ \citep[see 
Table~2 of][]{roy13a}. Our conclusions would remain essentially unchanged were 
we to use the optically-thin estimate instead of the isothermal one.

\section{Can all the detected \hii\ absorption originate in the CNM ?}
\label{sec:cnm}

One of the standard tenets of the ISM literature is that the warm neutral 
medium contributes to the \hii\ emission but is not detectable in \hii\ 
absorption due to its low optical depth \citep[e.g.][]{kulkarni88}. Wide 
absorption components are hence generally assumed to arise from non-thermally 
broadened CNM. For example, this is the model used by \citet{heiles03a} to 
model their \hii\ absorption/emission spectra. Since our survey was designed 
to have sufficient sensitivity to detect absorption from the WNM, the first 
question that one would like to address is: do our absorption spectra in fact 
definitely contain absorption from gas which is not in the CNM phase, or can 
all the absorption be accounted for by the CNM?

\begin{figure}
\begin{center}
\includegraphics[scale=0.4]{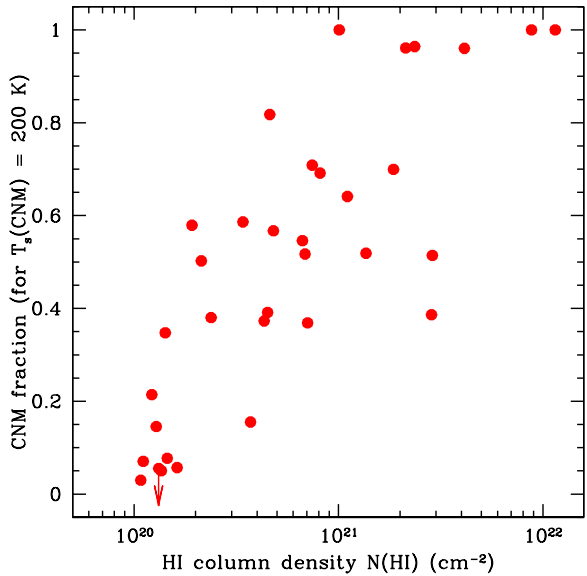}
\caption{\label{fig:cnm} The maximum CNM fraction along each sightline, assuming a CNM spin temperature of $200$\,K, plotted as a function of total \hi\ column density. See Section~\ref{sec:cnm} for details.}
\end{center}
\end{figure}

In order to address this issue, we consider a strawman model in which all the 
\hii\ absorption arises in the CNM, at kinetic temperatures $40-200$~K (and 
$\Ts = \Tk$). The CNM column density along each sightline is then given by 
\begin{equation}
\label{eqn:nhicnm}
{\rm N(CNM)} = 1.823 \times 10^{18} \times T_{\rm s,CNM} \times \int \tau {\rm dV} \:\:,
\end{equation}
where $T_{\rm s,CNM}$ is the harmonic mean spin temperature of the CNM along 
the sightline. Since the CNM spin temperature range is $40-200$~K, the 
harmonic mean CNM spin temperature must lie within this range, with a maximum 
possible value of $200$~K. Thus, an upper limit to the CNM column density for 
each sightline can be obtained by assuming $T_{\rm s,CNM} = 200$~K in 
equation~\ref{eqn:nhicnm}, and using the values for $\int \tau {\rm dV}$ from 
column~(3) of Table~\ref{table:sample}. The ratio of this upper limit to the 
total \hi\ column density listed in column~(2) of the table then gives an 
upper limit to the CNM fraction along the sightline. Fig.~\ref{fig:cnm} plots 
this upper limit to the CNM fraction on each sightline, i.e. the ratio 
N(CNM)/N(\hi), against the total \hi\ column density N(\hi). We obtain a 
median CNM fraction of $\sim 0.52$ in the sample of 34~sources, with values 
ranging from $< 0.055$ to $1$. It should be emphasized that these are {\it 
upper} limits to the CNM fraction, because we have assumed that (1)~all the 
observed absorption arises from the CNM and (ii)~the CNM has a harmonic mean 
spin temperature of 200~K, the highest value in the CNM range.

\begin{figure}
\begin{center}
\includegraphics[scale=0.4]{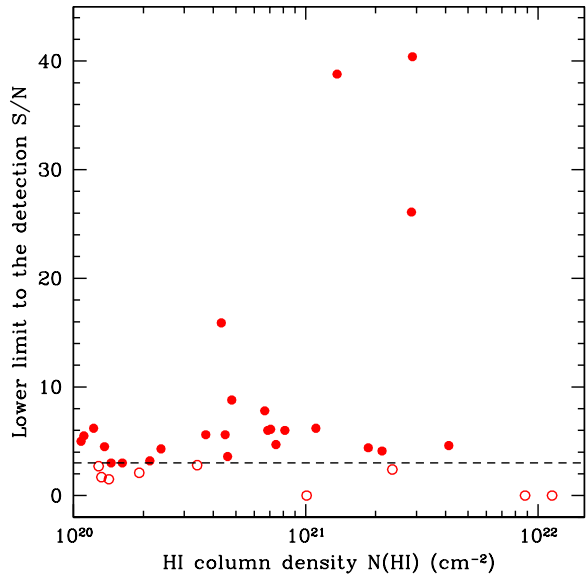}
\caption{\label{fig:wnm-snr} The S/N at which ``residual'' gas with $\Ts = 5000$\,K and line FWHM~$= \Delta V_{90}$ (of the LAB emission spectra) would have been detected in the \hii\ absorption spectra of our sample; the dashed horizontal line indicates a detection of $3\sigma$ significance. This assumes a simple model in which all of the detected \hii\ absorption arises from the CNM, with $\Ts = 200$\,K, while the emission arises in both CNM and WNM. The figure shows that \hii\ absorption from the residual gas would have been detected at high significance along 25 of the 34 sightlines of the full sample of Paper~I, even for very conservative assumptions. See Section~\ref{sec:cnm} for details.}
\end{center}
\end{figure}

Next, for each sightline, the difference [N(\hi) - N(CNM)] then gives the 
amount of ``residual'' \hi\ with $\Tk > 200$~K along each sightline. Note 
that, in our strawman model, this gas is {\it not detected in absorption}. We 
can thus immediately test the veracity of the model by checking whether the 
sensitivity in the individual spectra is sufficient to detect this residual 
gas in absorption. For this purpose, we assume a worst-case scenario, wherein 
the residual gas has $\Ts = 5000$~K and line FWHM equal to $\Delta V_{90}$ of 
the emission profile. These are extremely conservative assumptions, as both 
the high spin temperature and the large line FWHM reduce the peak \hii\ 
optical depth, making it more difficult to detect a given \hi\ column density 
in absorption. With these assumptions, we compute \dnhi, the $1\sigma$ 
sensitivity of each spectrum to \hii\ absorption from gas with the above line 
FWHM and spin temperature. 

\setcounter{figure}{2}
\begin{figure*}
\begin{center}
\includegraphics[width=3.5in,angle=-90]{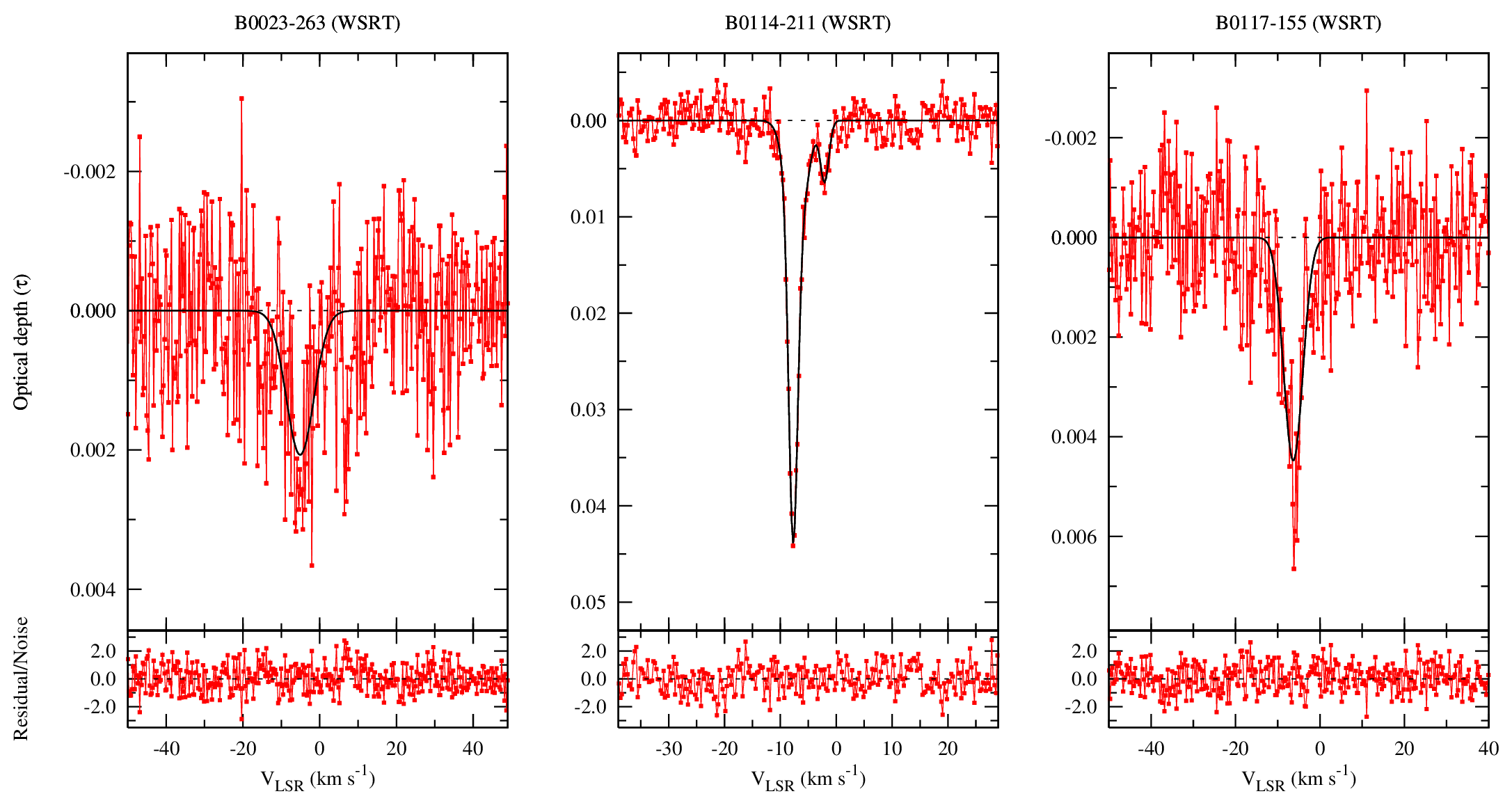}
\includegraphics[width=3.5in,angle=-90]{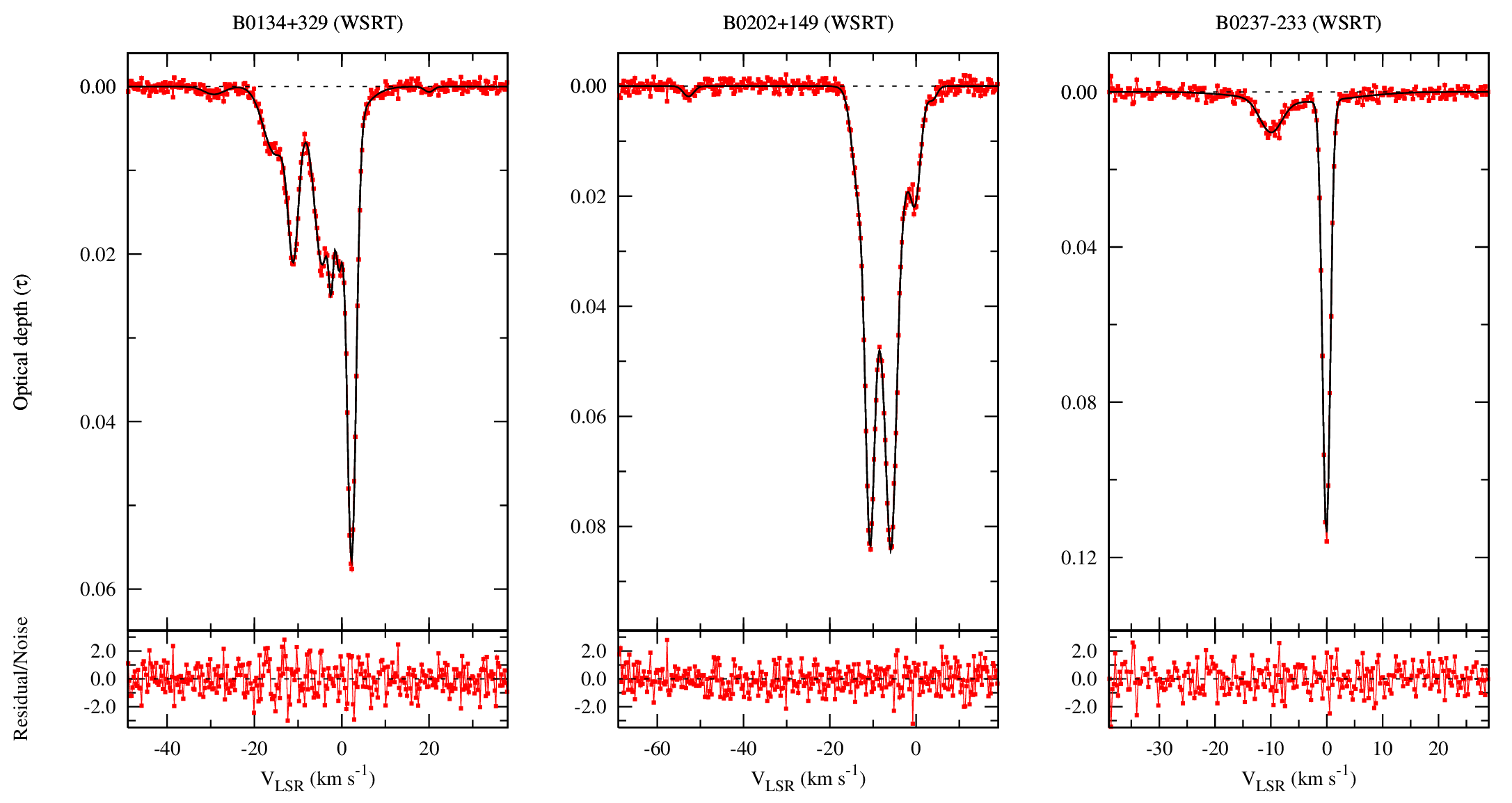}
\caption{\label{fig:spectra}\hii\ absorption spectra and multi-Gaussian fits for the 30 sources of the sample, ordered by right ascension. The source name and the telescope used to obtain the spectrum is listed above each figure. For each source, the upper panel contains the multi-Gaussian fit overlaid on the \hii\ optical depth profile, while the lower panel shows the fit residuals (normalized by the RMS optical depth noise). In most cases, the full velocity range covered by the absorption spectrum is not shown, to make it easier to view the profile.}
\end{center}
\end{figure*}

\setcounter{figure}{2}
\begin{figure*}
\begin{center}
\includegraphics[width=3.5in,angle=-90]{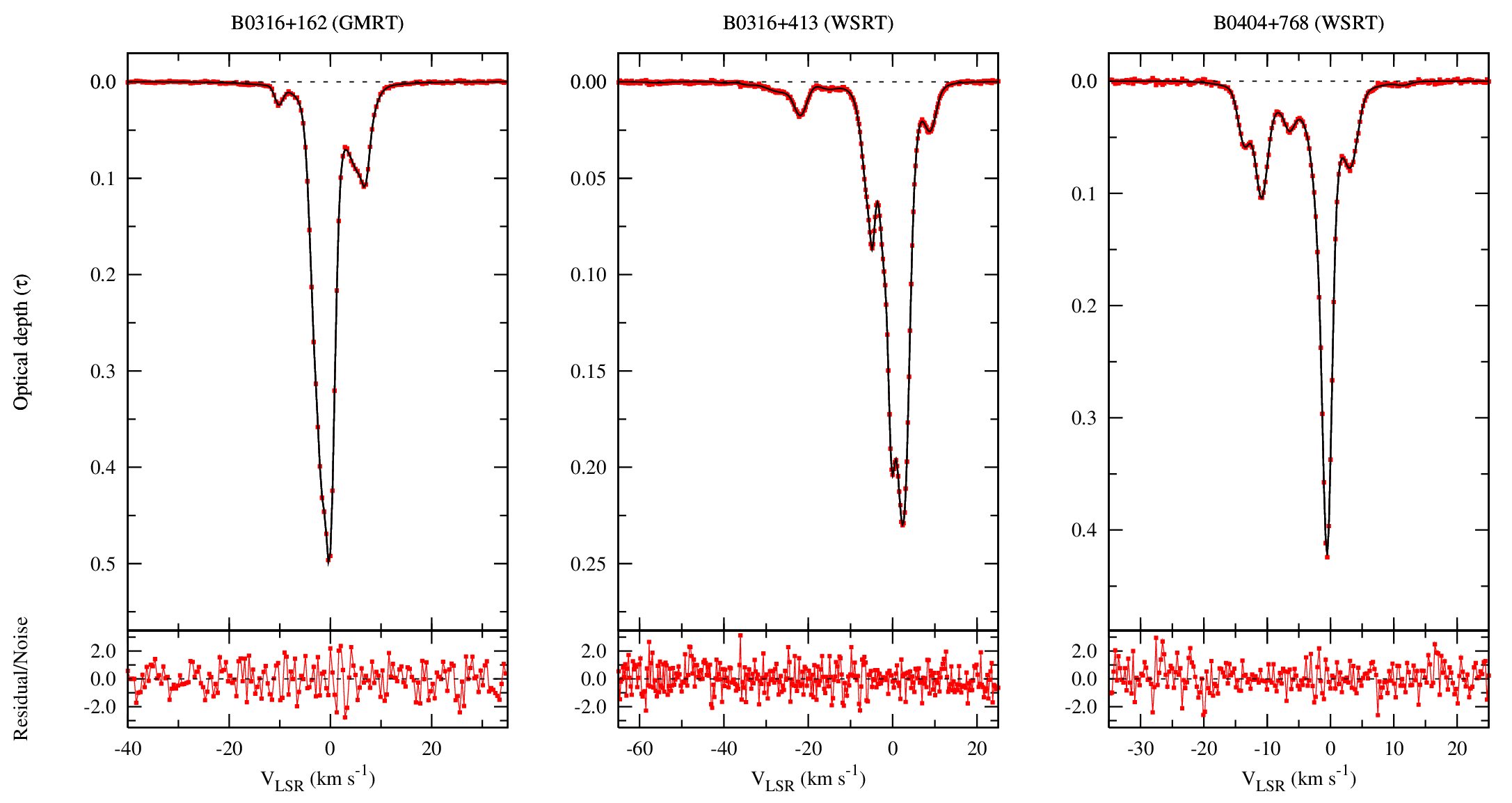}
\includegraphics[width=3.5in,angle=-90]{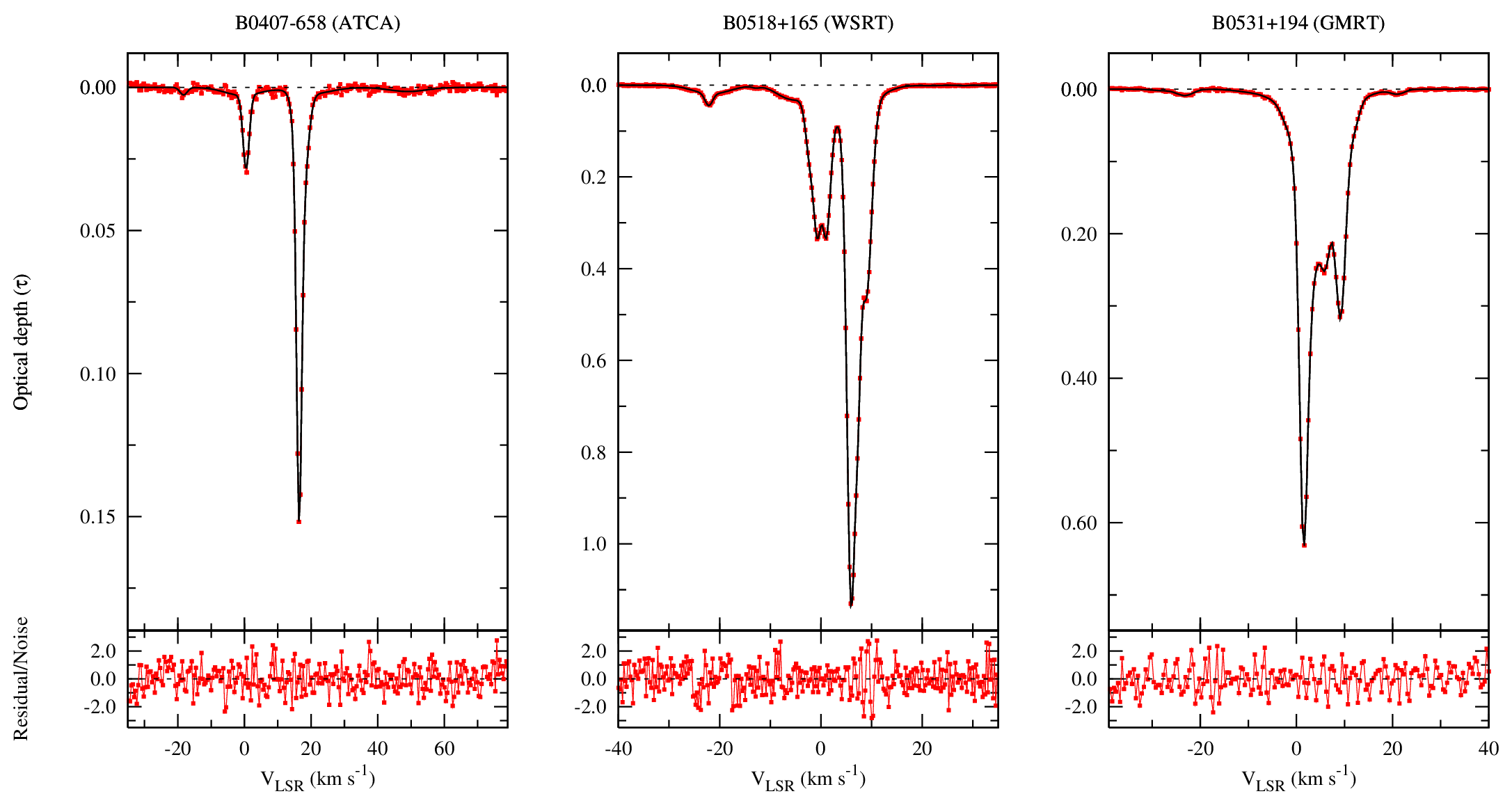}
\caption{(contd.) \hii\ absorption spectra and multi-Gaussian fits for the 30 sources of the sample.}
\end{center}
\end{figure*}

\setcounter{figure}{2}
\begin{figure*}
\begin{center}
\includegraphics[width=3.5in,angle=-90]{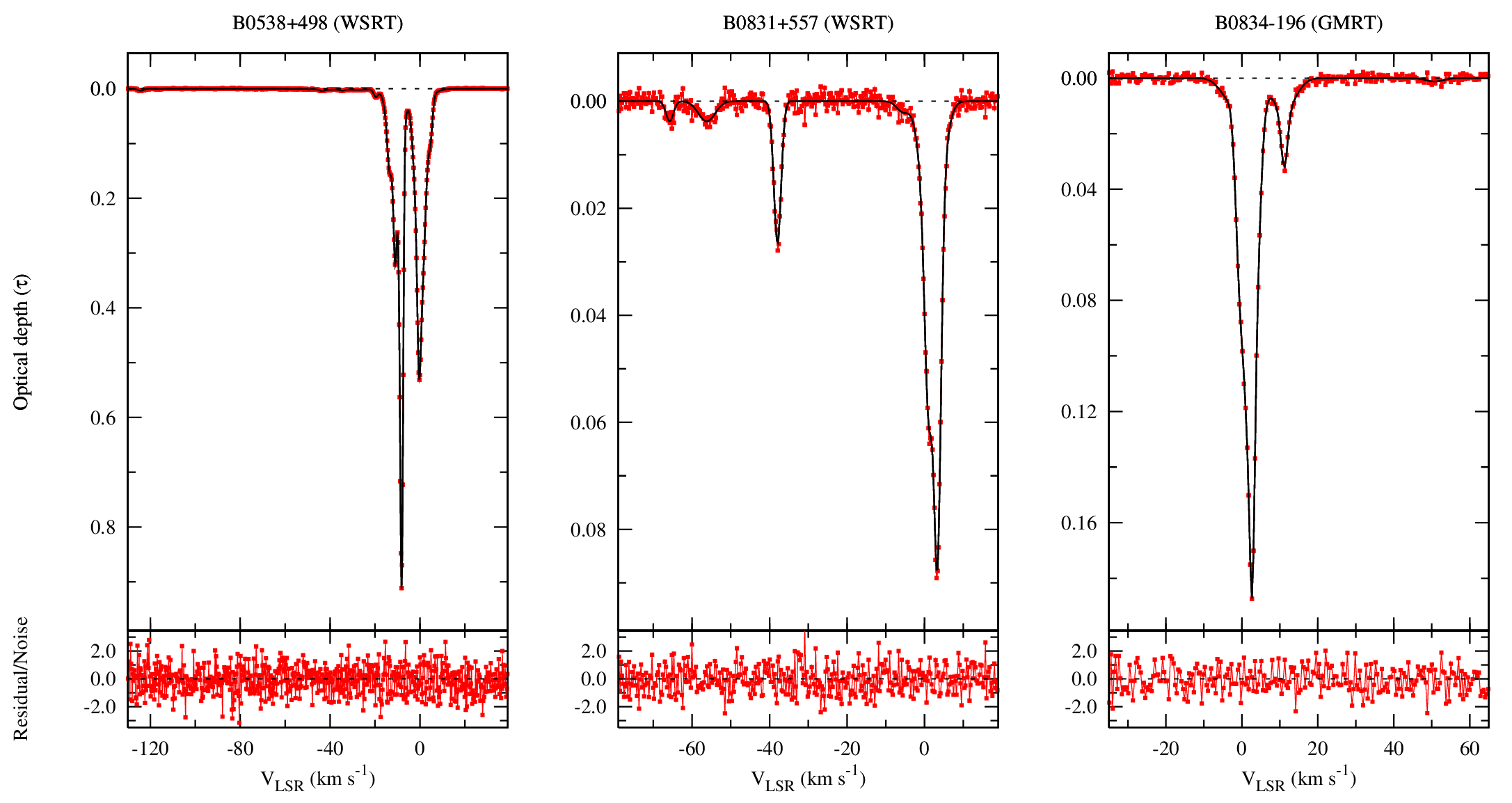}
\includegraphics[width=3.5in,angle=-90]{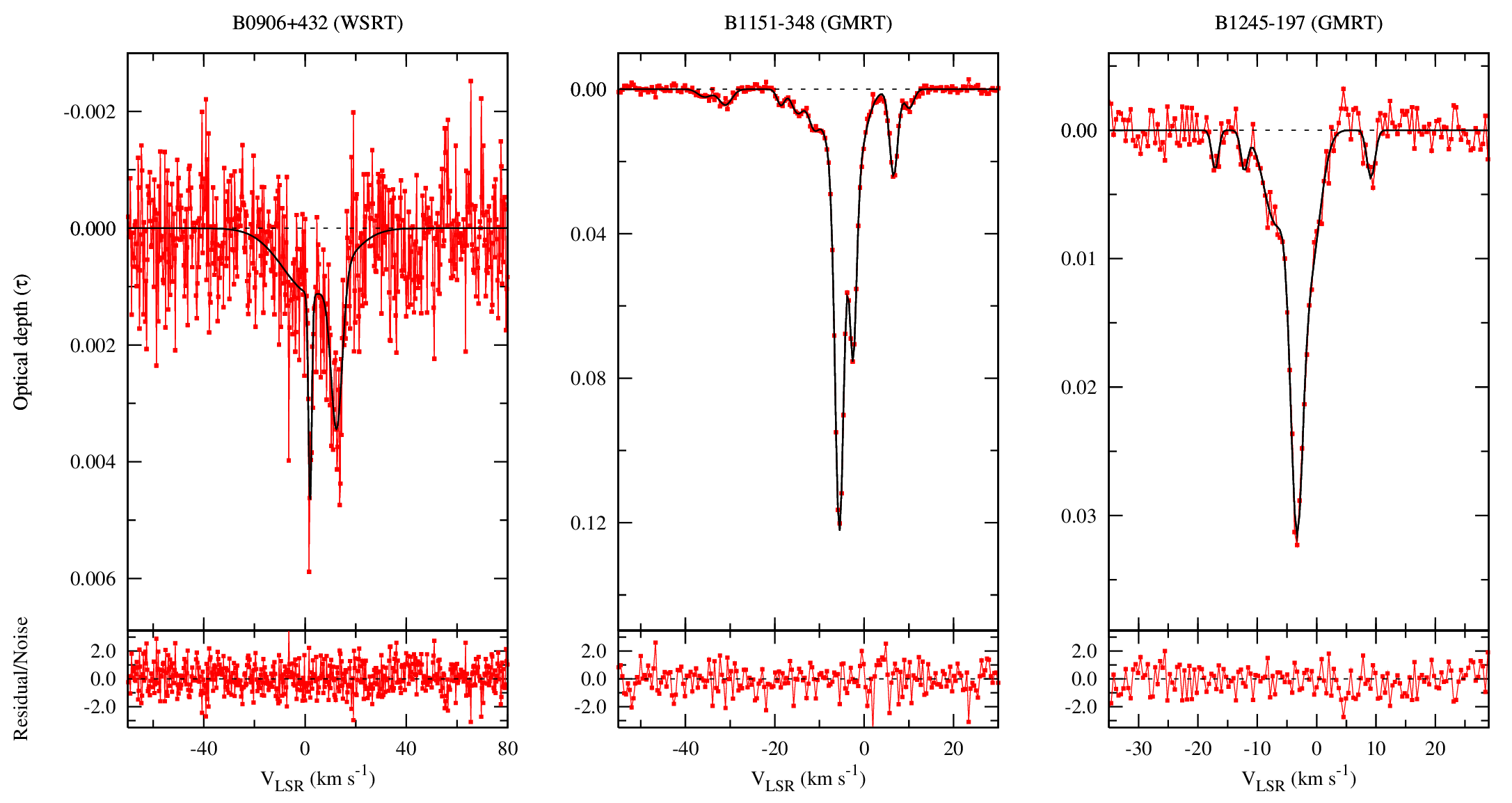}
\caption{(contd.) \hii\ absorption spectra and multi-Gaussian fits for the 30 sources of the sample.}
\end{center}
\end{figure*}

\setcounter{figure}{2}
\begin{figure*}
\begin{center}
\includegraphics[width=3.5in,angle=-90]{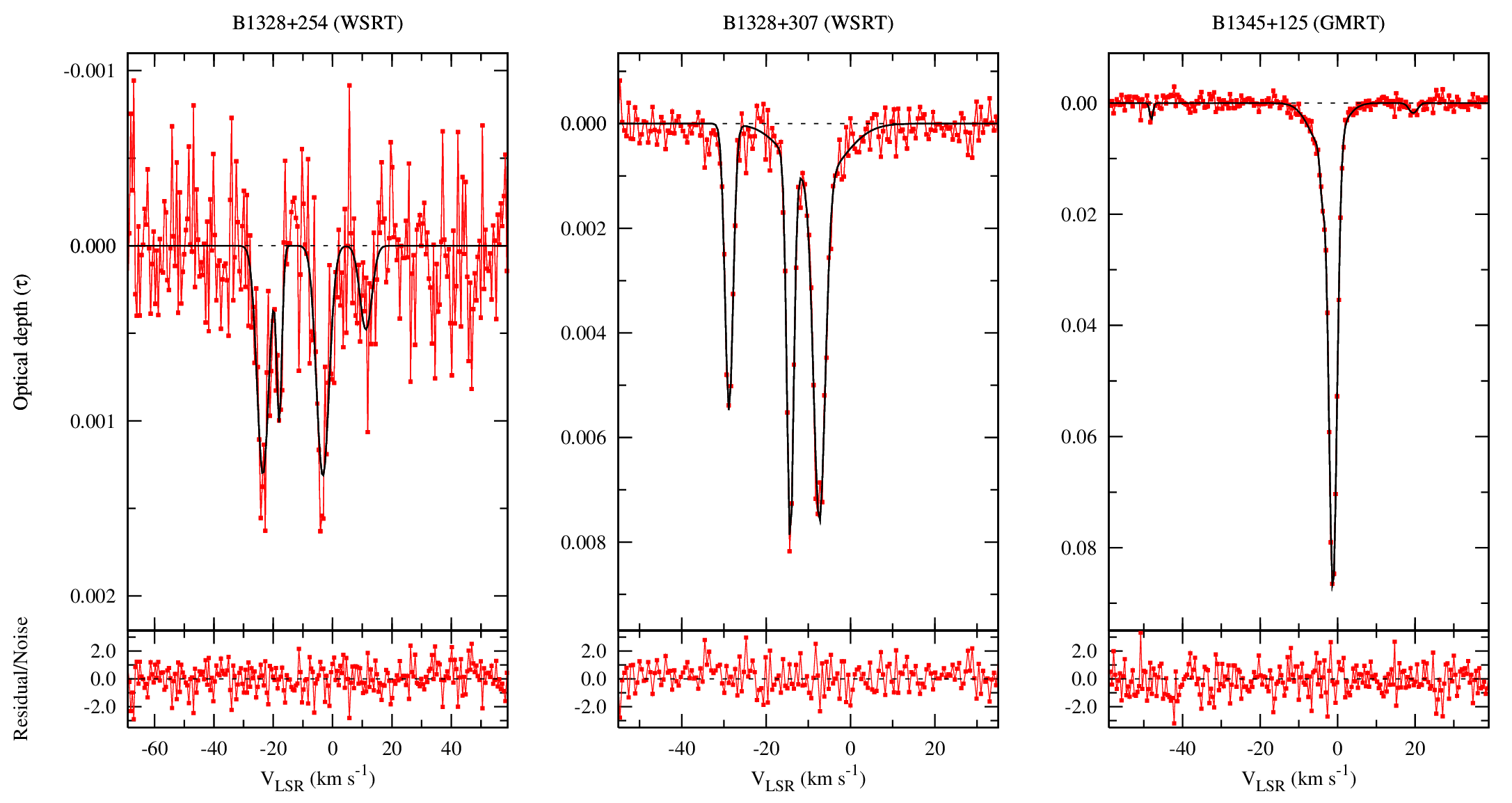}
\includegraphics[width=3.5in,angle=-90]{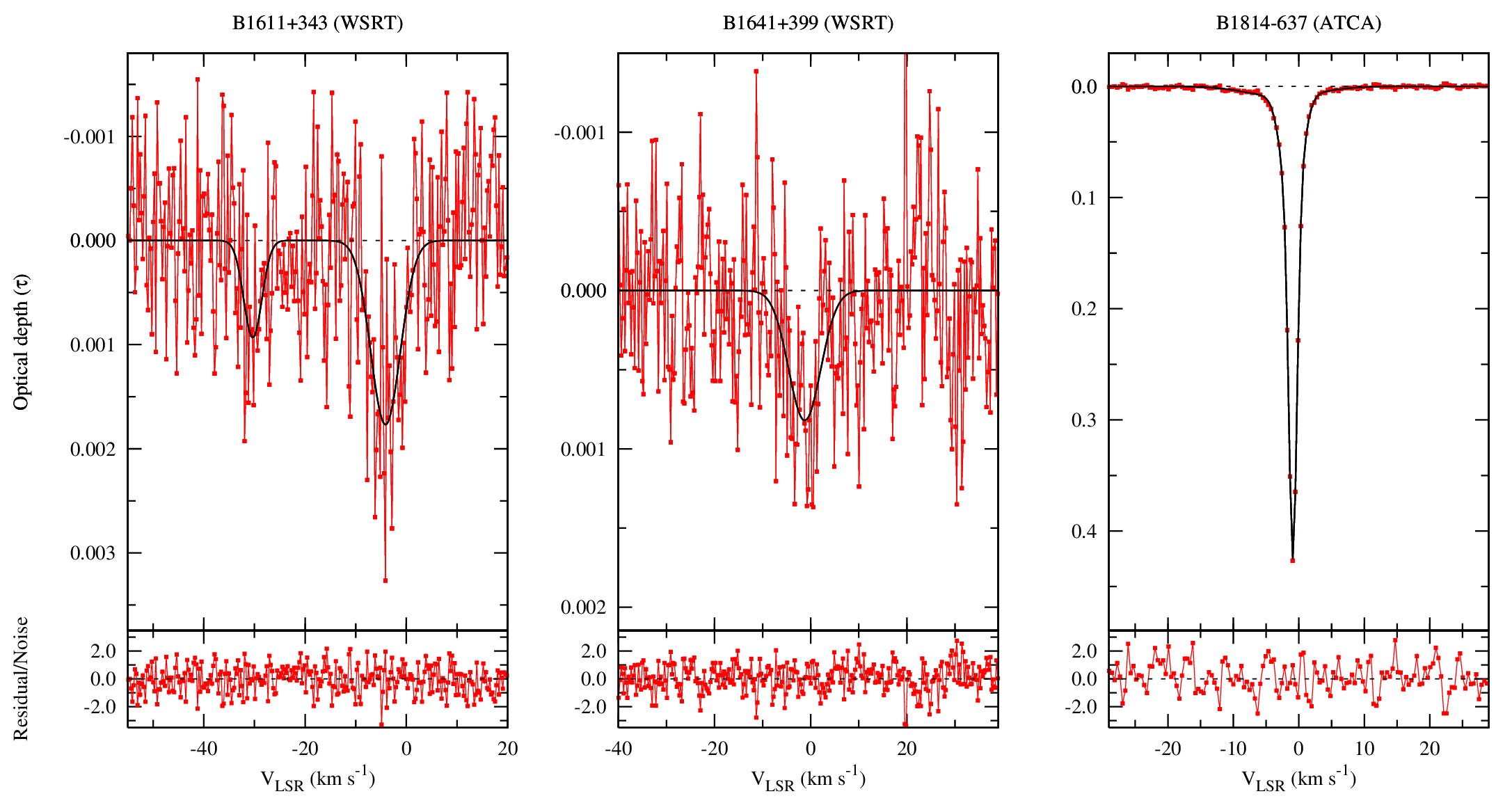}
\caption{(contd.) \hii\ absorption spectra and multi-Gaussian fits for the 30 sources of the sample.}
\end{center}
\end{figure*}

\setcounter{figure}{2}
\begin{figure*}
\begin{center}
\includegraphics[width=3.5in,angle=-90]{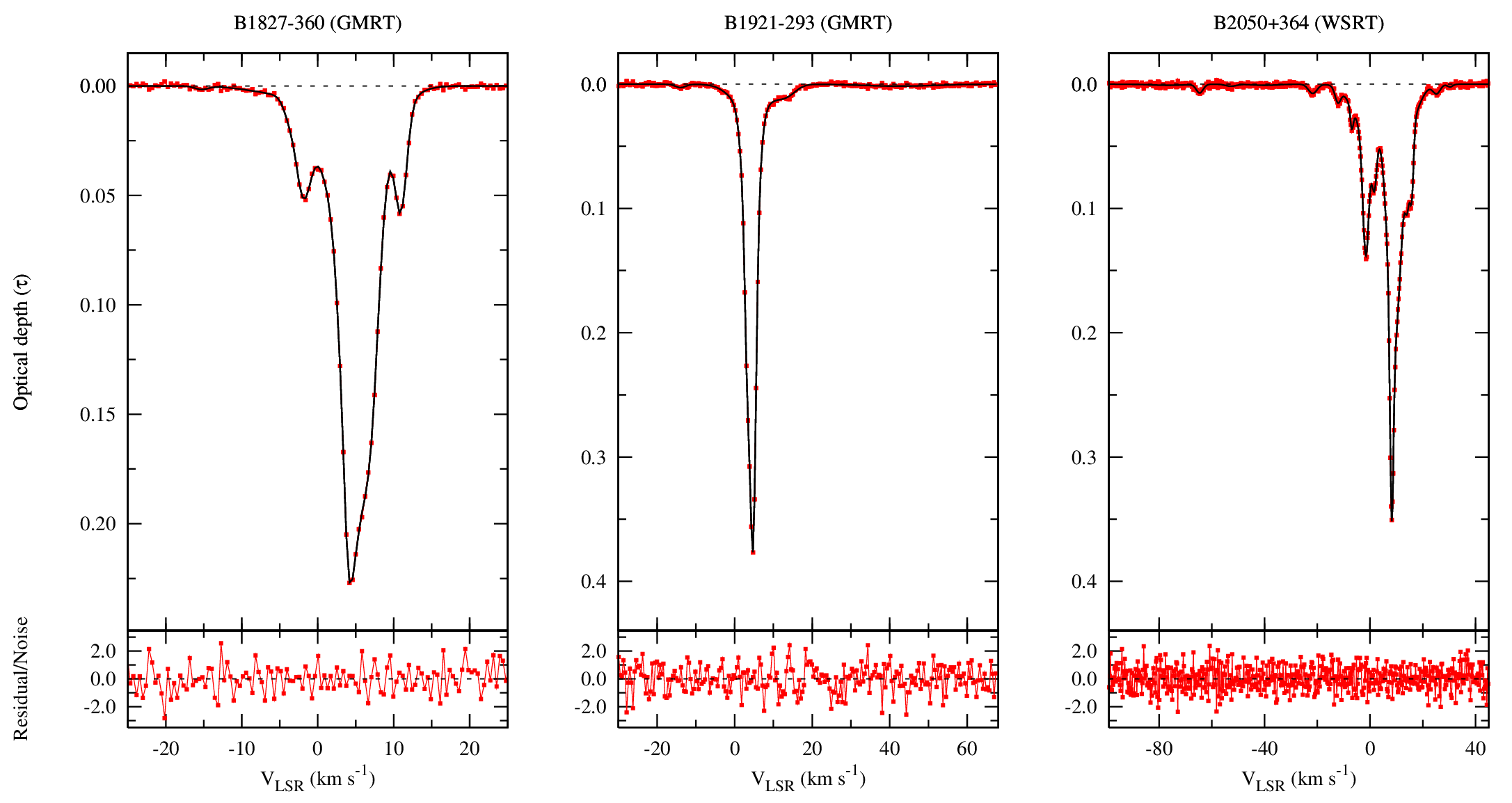}
\includegraphics[width=3.5in,angle=-90]{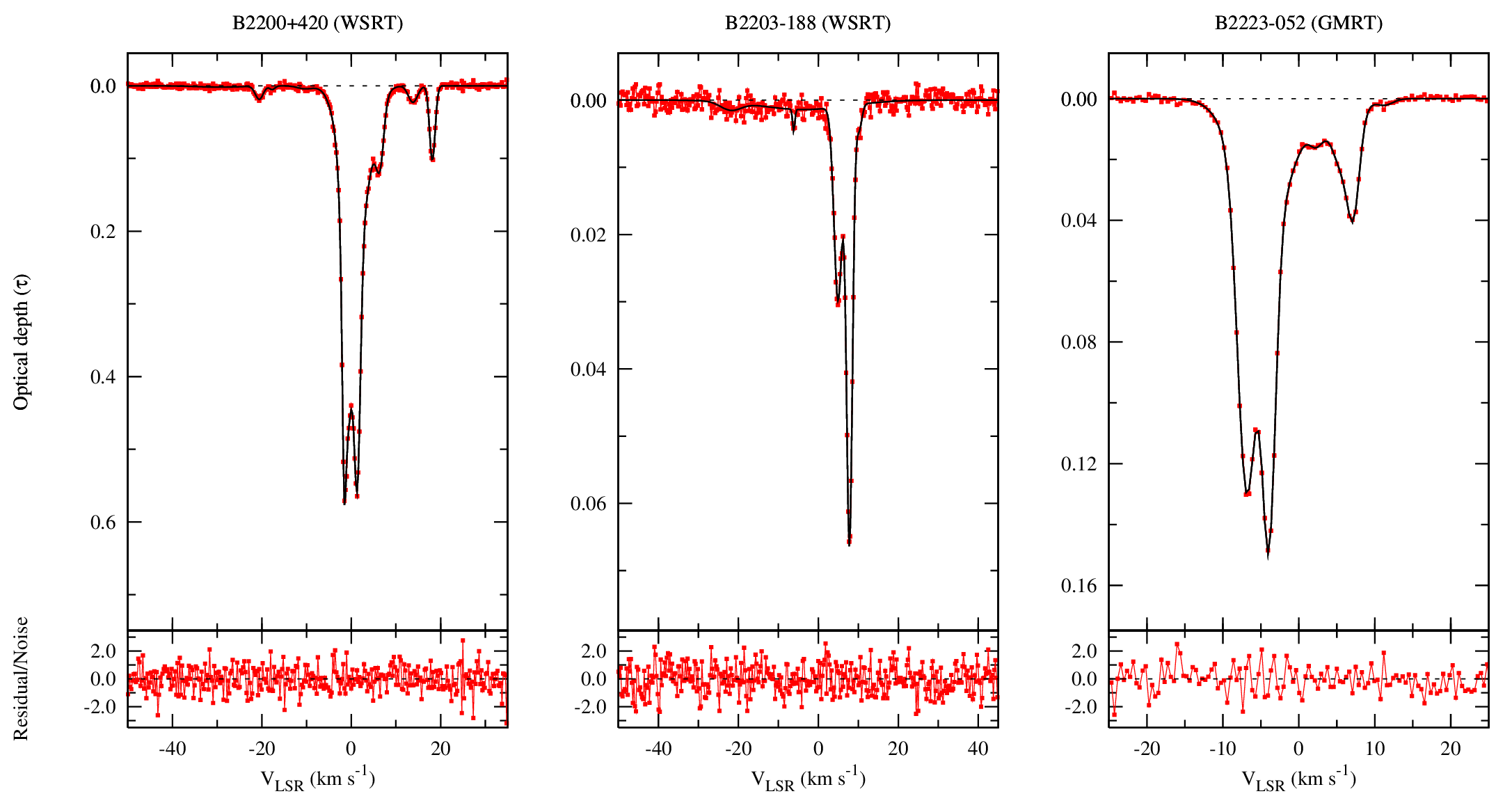}
\caption{(contd.) \hii\ absorption spectra and multi-Gaussian fits for the 30 sources of the sample.}
\end{center}
\end{figure*}

\begin{figure*}
\begin{center}
\includegraphics[width=3.5in,angle=-90]{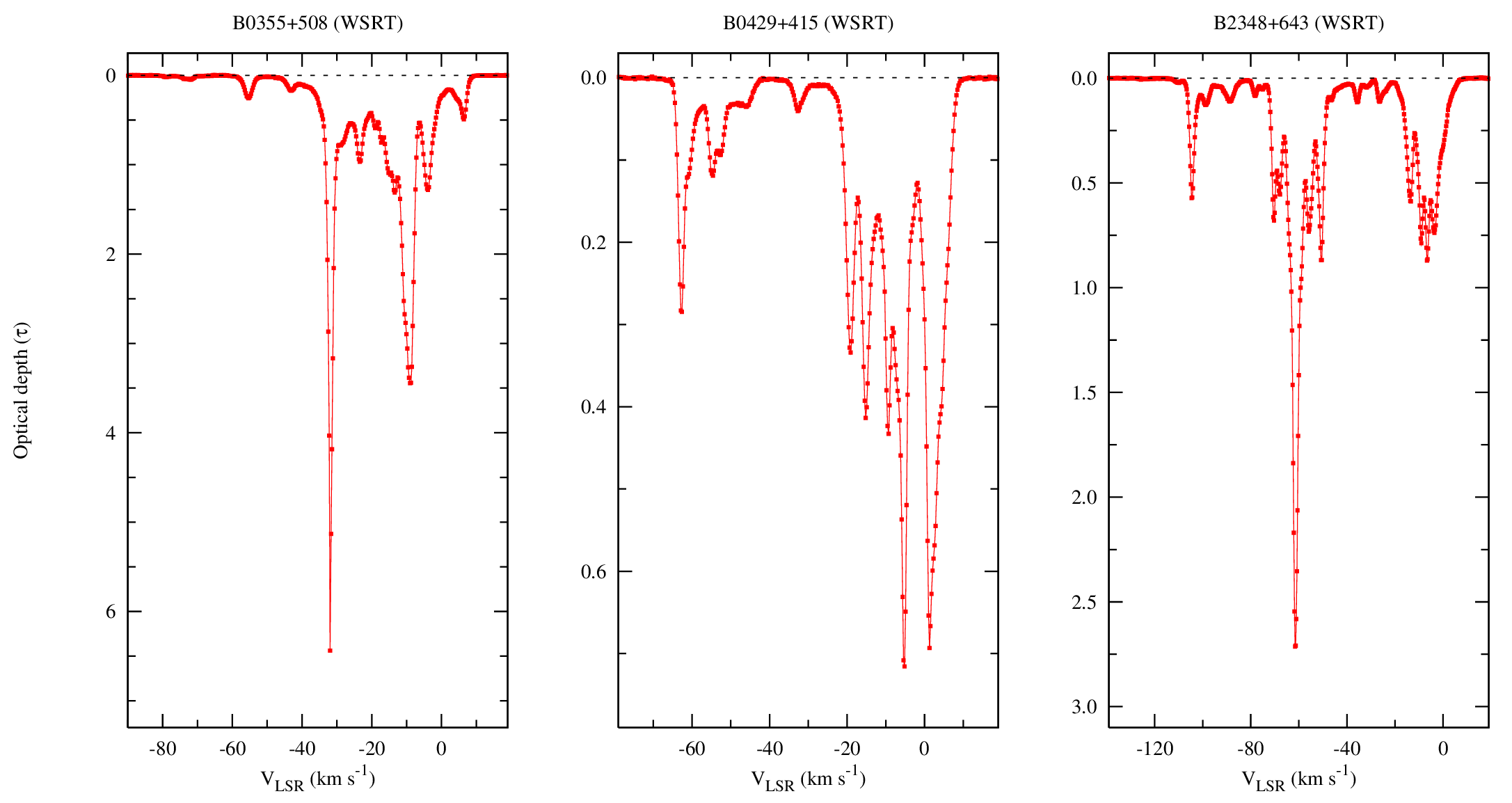}
\caption{\label{fig:spectra5}\hii\ absorption spectra for the three sources (B0355+508, B0429+415 and B2348+643), whose \hii\ profiles were found to be too complex to obtain a reliable fit.}
\end{center}
\end{figure*}

The results of this analysis are summarized in Fig.~\ref{fig:wnm-snr}, which 
plots the signal-to-noise ratio (S/N) at which absorption from the residual 
gas would have been detected in our \hii\ absorption spectra versus the \hi\ 
column density; the dashed horizontal line indicates a $3\sigma$ detection. 
There are three sightlines (towards B0355$+$508 and B2348$+$643, the two 
sources with the lowest Galactic latitudes, and B0316$+$362) where the (S/N) 
is listed as zero, as the derived CNM column density is larger than the total 
\hi\ column density. For these sightlines, the average CNM spin temperature 
must be lower than the assumed 200\,K, and one cannot rule out the possibility 
that all the \hii\ absorption arises in the CNM (or, indeed, that all the \hi\ 
along the sightline is in the CNM!). However, in 25 of the 34 cases, the 
spectra are sufficiently sensitive to detect \hii\ absorption from the 
residual gas (at $\ge 3\sigma$ significance), even for our very conservative 
assumptions. We emphasize that gas at lower spin temperatures ($\Ts < 5000$~K) 
or with line FWHMs lower than $\Delta V_{90}$ (as would be typical) would have 
been detected at even higher significance. We thus find that a strawman model 
in which all the absorption arises from the CNM is clearly ruled out: {\it 
most of our \hii\ absorption spectra must contain absorption from gas with 
$\Ts >> 200$\,K}. We will return to the issue of the actual temperature of the 
absorbing gas in Sec.~\ref{sec:2-phase}.

\section{The Gaussian parametrisation}
\label{sec:gauss}

The simplest way of modelling an absorption profile is to treat it as arising 
from a number of absorbing ``clouds'' along the line of sight. If each cloud 
is individually in equilibrium, at a kinetic temperature $\Tk$, its line 
profile would be a ``Voigt'' profile, a convolution of a Gaussian function 
with a Lorentzian, which reduces to a Gaussian for unsaturated absorption 
lines. As such, the simplest physically-motivated model for an unsaturated 
absorption profile consists of a sum of multiple Gaussian components, with the 
width of each component determined by the kinetic temperature of the gas. In 
principle, such a Gaussian decomposition allows one to derive the kinetic 
temperature of individual ``clouds''. Modelling absorption spectra as a 
superposition of Gaussian components has hence long been popular in \hii\ 
absorption studies 
\citep[e.g.][]{mebold72,rad72,mebold82,lane00,kanekar01b,heiles03a,heiles03b,kanekar03b}.
Indeed, the results of \citet{heiles03a} that a significant fraction of the 
WNM is in the unstable phase are critically dependent on the use of such a 
multi-Gaussian model for the \hii\ emission and absorption spectra {\it and} 
the assumption that the line widths predominantly arise from thermal motions.

Unfortunately, it has also long been appreciated that the above interpretation 
is not necessarily physically meaningful \citep[e.g.][]{mebold72}. For 
example, it is not at all obvious that the individual Gaussian components 
correspond to distinct physical entities, i.e. whether a typical sightline 
consists of multiple clouds in internal equilibrium. Non-thermal motions are 
well known in galaxies, and could lead to distortions of the absorption 
profile from a Gaussian shape. Even in the case of turbulent broadening, which 
also yields a Gaussian profile, the final line width is determined by a 
combination of the kinetic temperature and the turbulent motions, and hence 
only yields an upper limit to the kinetic temperature. Finally, Gaussian 
functions do not form an orthogonal basis, implying that a decomposition is 
formally not unique, especially in the presence of noise. 

Attempts have also been made to model line profiles by assuming a specific 
form for the density and/or temperature distribution in \hi\ clouds and local 
isobaricity \citep[e.g.][]{braun05}. Note that theoretical models of the ISM 
are based on local isobaricity, as assumed in the above model, and do not 
require that individual clouds be isothermal. Conversely, the Gaussian 
decomposition is strictly valid for cases where each ``cloud'' along the 
sightline is isothermal. However, while it is certainly possible to model 
spectral lines along the lines of \citet{braun05}, there has so far been no 
obvious physical motivation for a specific functional form. It is probably 
fair to say that, at present, these approaches are less physically motivated 
than a Gaussian decomposition. Finally, while much progress has been made in 
hydrodynamic simulations of the ISM 
\citep[e.g.][]{gazol01,audit05,semadeni06,hennebelle07a,audit10,saury13}, the 
complexity of the problem has meant that such studies have unfortunately not 
yielded much information on realistic models of interstellar gas clouds.

In summary, while there are certainly possible drawbacks to using Gaussian 
components to model \hii\ spectra, there is no obvious alternative route to 
determining physical conditions in the ISM. Further, given the problems 
associated with single-dish \hii\ emission spectra \citep[e.g.][]{kanekar03b}, 
it is important to apply a multi-Gaussian component model to the ``cleaner'' 
\hii\ absorption profiles to test whether these too yield a large fraction of 
unstable atomic gas in the ISM. This is the broad approach we follow in this 
paper, using a Gaussian decomposition to determine the distribution of \hi\ 
kinetic temperatures, and to test whether a significant fraction of the 
neutral ISM is in the unstable phase. 

\subsection{The multi-Gaussian fits}
\label{sec:fit}

\begin{figure*}
\begin{center}
\includegraphics[scale=0.4]{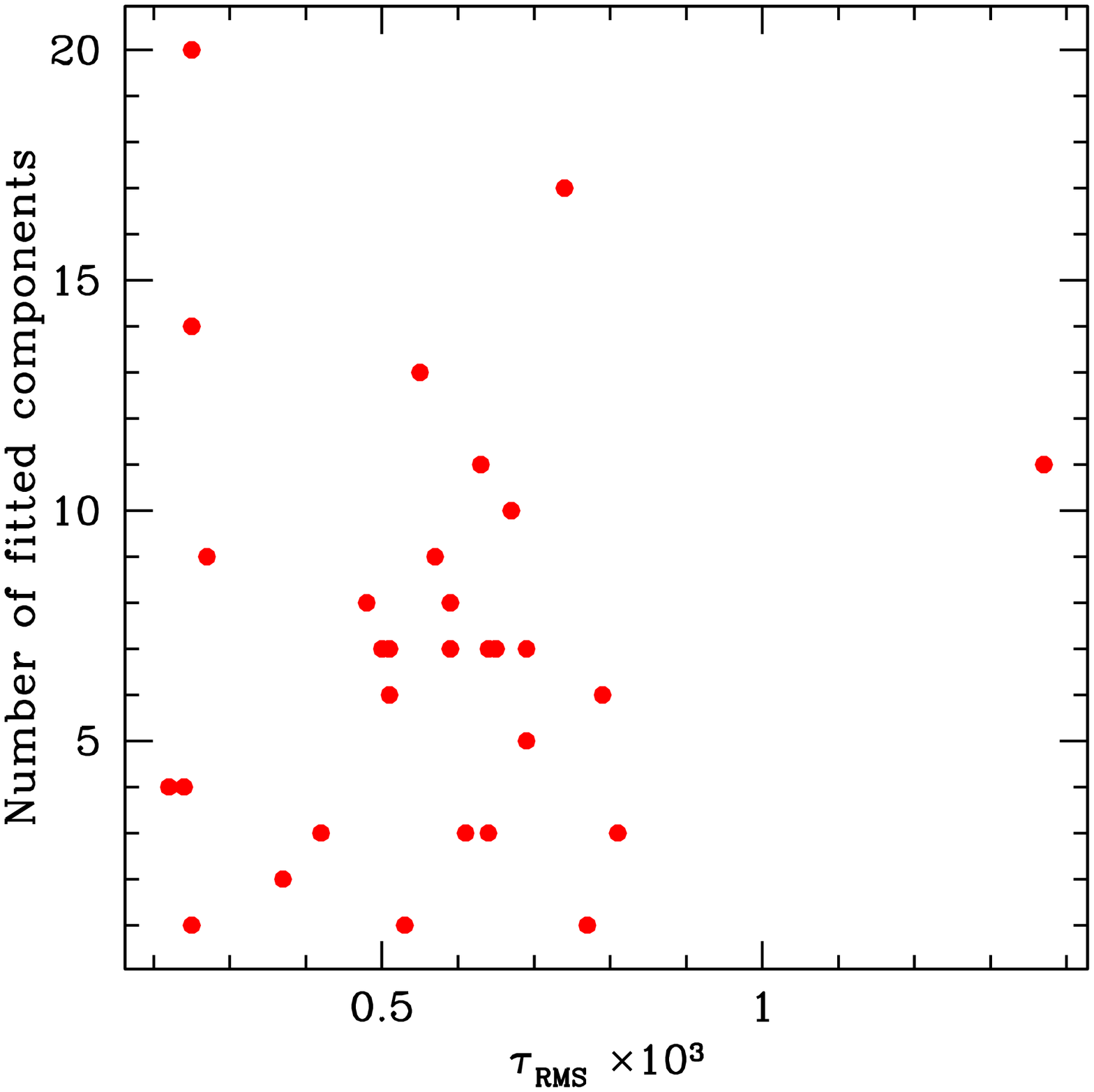}
\includegraphics[scale=0.4]{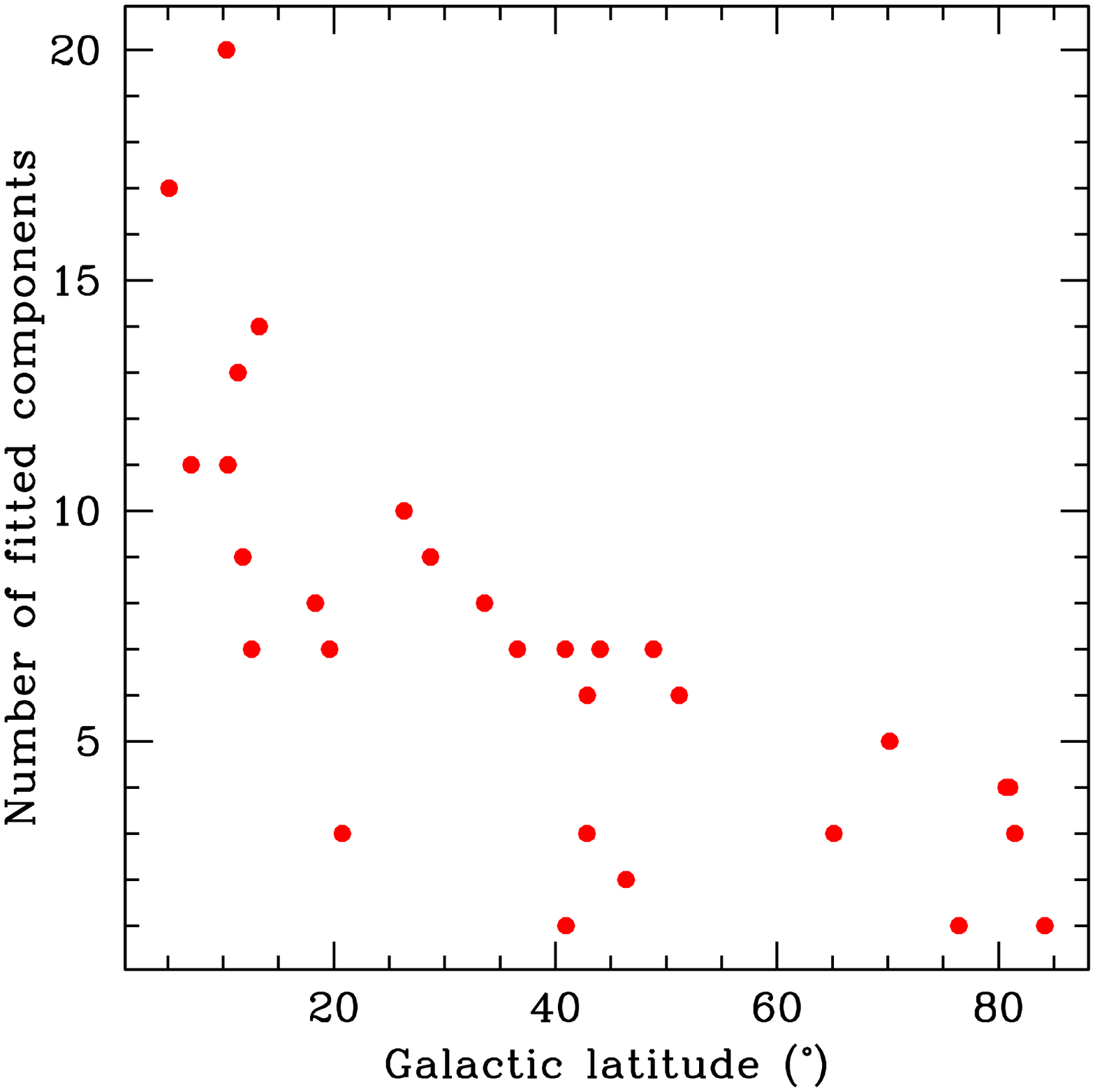}
\caption{\label{fig:ncomp} The number of fitted Gaussian components plotted 
versus [A]~(left panel) the off-line RMS optical depth noise of each spectrum 
and [B]~(right panel) the absolute Galactic latitude. See text for discussion.}
\end{center}
\end{figure*}

\begin{figure*}
\begin{center}
\includegraphics[scale=0.4]{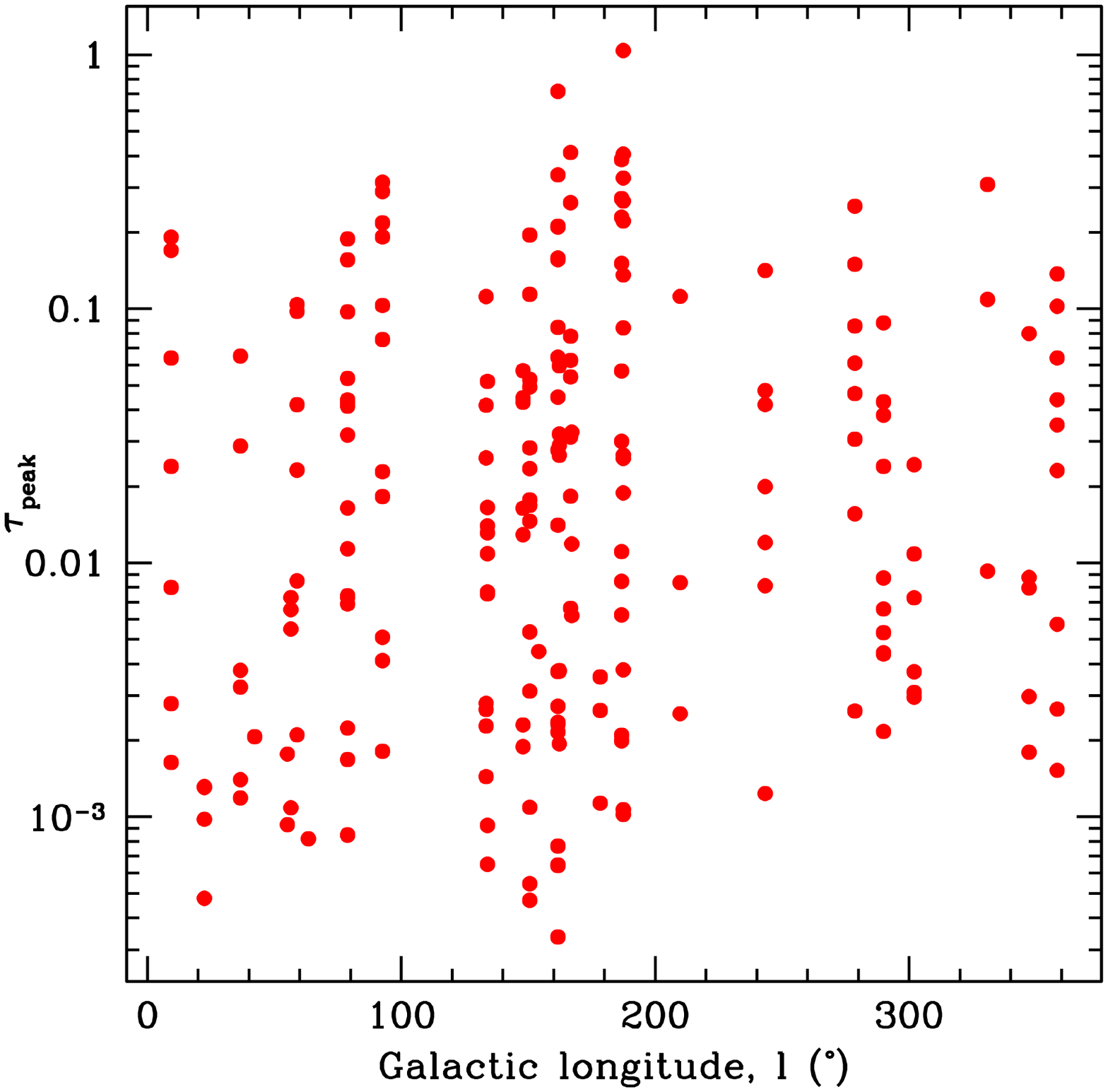}\includegraphics[scale=0.4]{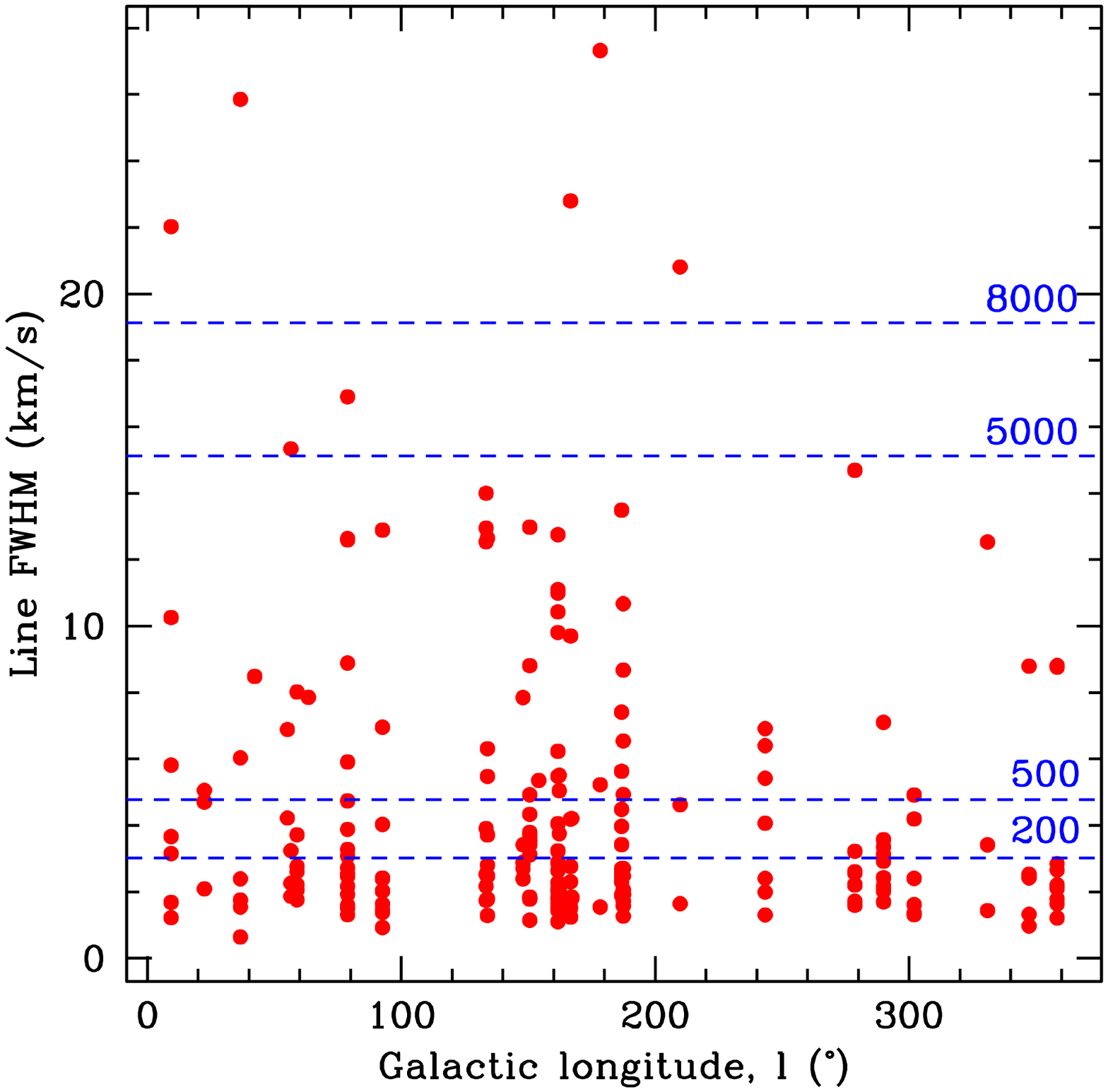}
\includegraphics[scale=0.4]{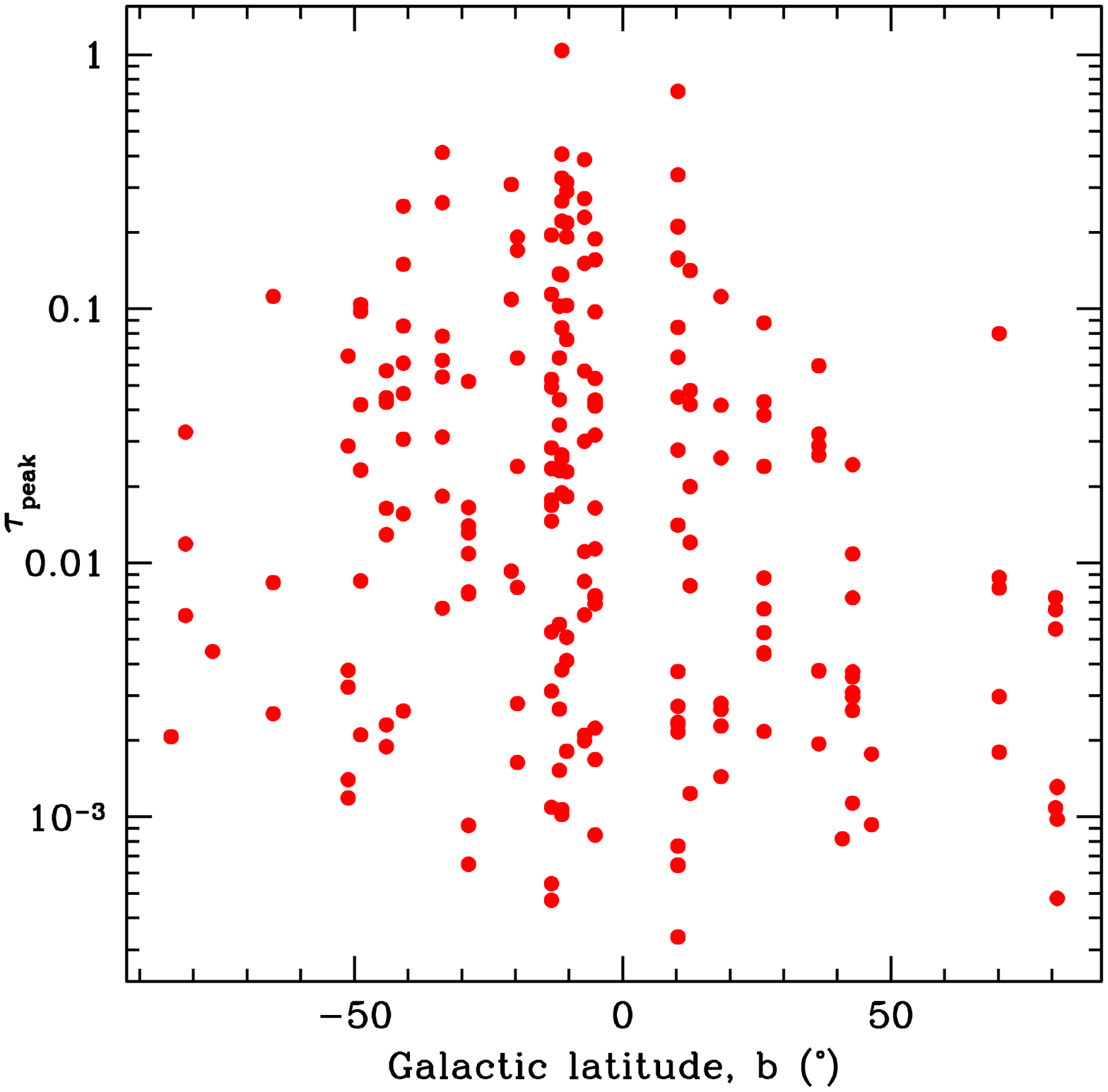}\includegraphics[scale=0.4]{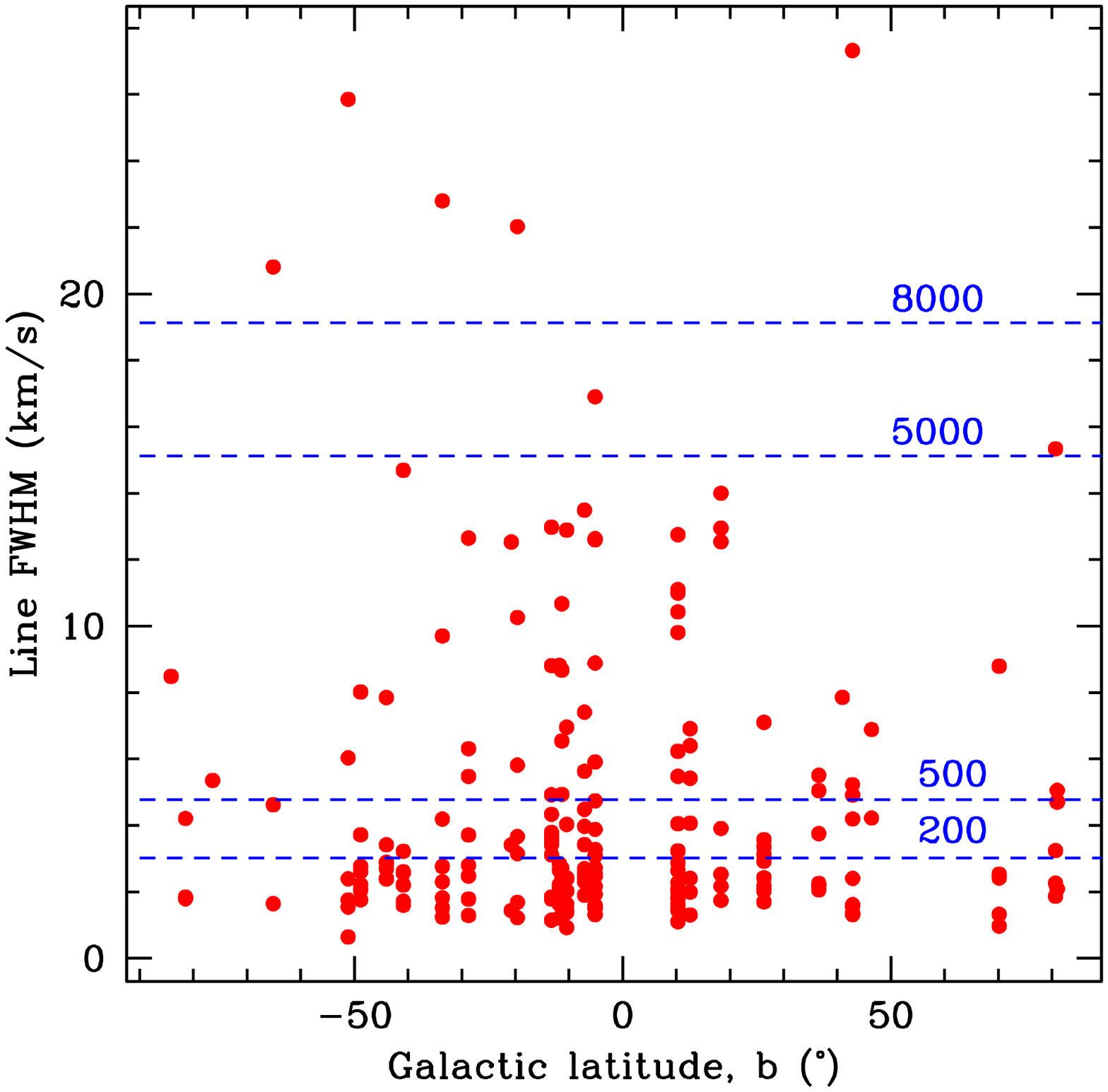}
\caption{\label{fig:tvlb} Distribution of component parameters (peak optical 
depth and velocity FWHM) with Galactic longitude and latitude. The horizontal 
dashed lines on the left panels indicate the velocity FWHMs corresponding to 
kinetic temperatures $T_{\rm K} = 200, 500, 5000$ and $8000$\,K. See text 
for discussion.}
\end{center}
\end{figure*}

As discussed in Paper~I (see Section~2.3), the noise is not uniform across 
each absorption spectrum, but is higher at velocity ranges containing Galactic 
\hii\ emission, which contributes to the system temperature in these velocity 
channels. This was taken into account to derive optical depth noise spectra 
for each target source. We then carried out an iterative multi-Gaussian fit to 
each of the 33 optical depth spectra with detected \hii\ absorption (i.e. 
excluding the sightline towards B0438$-$436) using the noise spectrum for the 
sightline, and the standard technique of $\chi^2$-minimization, using the 
Levenberg-Marquardt algorithm. For each spectrum, the number of Gaussian 
components was increased (beginning with a single component) until the reduced 
chi-square ($\chi_r^2$) was as close to unity as possible, with noise-like 
residuals. For three sources (B0355$+$508, B0429$+$415 and B2348$+$643), the 
absorption profiles are extremely complex and impossible to fit with a 
reasonable number ($\lesssim 20$) of spectral components. We have hence 
dropped these three spectra from the multi-component analysis, thus retaining 
30~sources in the final sample of sources for which the multi-Gaussian fits 
were found to be successful.

For each sightline, we verified through a Kolmogorov-Smirnov rank-1 test and 
an Anderson-Darling test that the residual spectrum after subtracting out the 
fitted model was consistent with a normal distribution (within $3\sigma$ 
significance in each test). Since the noise is channel-dependent, these tests 
were carried out after scaling the residual spectrum by the noise spectrum 
(i.e. on the residual spectrum in S/N units). For all fits, we obtained 
$\chi_r^2 \approx 1$ (with $0.76 < \chi_r^2 < 1.29$ in all cases), indicating 
that the fitted model is a good representation of the observed \hii\ optical 
depth spectrum. Following standard procedures to obtain  error estimates, the 
uncertainties of the fit parameters were derived after scaling the errors so 
as to obtain $\chi_r^2 = 1$. The number of fitted components in each spectrum 
ranges from 1 (for B0023$-$263, B0117$-$155 and B1641$+$399) to 20 (for 
B0538$+$498), with a median number of 7.

The above fitting procedure yielded estimates for the peak optical depths, 
central velocities and line FWHMs ($\Delta V$) for the different Gaussian 
components. The doppler temperature $\Td$ of each component was then obtained 
from the expression
\begin{equation}
\Td = 21.855 \times \Delta V^2 \:\: .
\end{equation}
We emphasize that this is an {\it upper limit} to the kinetic temperature of 
each component.

For each of the 30~sightlines, the fitted multi-Gaussian model is overlaid on 
the optical depth spectrum in the upper panel of each plot in 
Fig.~\ref{fig:spectra}, while the lower panel of each plot shows the residual 
(S/N) spectrum (after dividing the residuals by the RMS optical depth noise). 
Fig.~\ref{fig:spectra5} shows the \hii\ absorption profiles of the three 
sources (B0355$+$508, B0429$+$415 and B2348$+$643) whose spectra were found to 
be too complex to obtain a reliable fit. Finally, Table~A1 in the Appendix 
contains the fit parameters and errors, the $\chi_r^2$ values and the derived 
doppler temperatures and errors. Note that the errors on the doppler 
temperatures have been corrected for the minor deviations from $\chi_r^2$ from 
unity.

It is interesting to ask whether the fitted multi-Gaussian profiles provide a 
realistic model of conditions in the ISM (e.g. whether each Gaussian component 
represents an \hi\ ``cloud'') or whether they merely provide a tool by which 
to fit a complicated profile. In the latter situation, one would expect that 
the number of Gaussians needed to fit a profile would increase with decreasing 
RMS optical depth noise, as more and more complex structure would become 
apparent with improved sensitivity. One would then expect an anti-correlation 
between the number of fitted components and the off-line RMS optical depth 
noise. Fig.~\ref{fig:ncomp}[A] plots the number of fitted components against 
the off-line RMS optical depth noise (at a uniform velocity resolution of 
$1$~\kms) for the 30 sightlines of the sample. It is clear from the figure 
that the number of components is not related to the RMS optical depth noise: 
for example, the two most sensitive spectra, towards B1328$+$254 and 
B1328$+$307 (with $\tau_{\rm RMS} \approx 0.00022-0.00024$ per 1~\kms\ 
channel), require only 4 Gaussian components, while the spectrum towards 
B1641$+$399 (with $\tau_{\rm RMS} \approx 0.00025$ per 1~\kms\ channel) 
requires a single Gaussian component for a good fit. 

It also appears plausible that low-latitude sightlines would traverse more 
\hi\ clouds, while high-latitude sightlines would only intersect a few clouds. 
One would then expect more components along low-latitude sightlines and fewer 
components along high latitude ones, independent of the sensitivity of the 
spectra. Fig.~\ref{fig:ncomp}[B] plots the number of components versus the 
absolute Galactic latitude; it is clear from the figure that far more 
components are required to obtain a good fit for low latitudes, $b \lesssim 
15^\circ$, than at high latitudes. Specifically, all sightlines with $b \geq 
15^\circ$ are well fitted with $\leq 10$ components, while all sightlines with 
$b \geq 40^\circ$ are well fitted with $\leq 7$ components. Conversely, all 
sightlines with $b \lesssim 10^\circ$ require $> 10$ components to obtain a 
good fit. We note, in passing, that the three sightlines which were excluded 
from the sample due to profile complexity were all at low latitudes, $b < 
10^\circ$.

Both the above results suggest that the fitted multi-Gaussian profiles may 
provide a realistic model for the sightlines in our sample. We note, however, 
that this may break down for the most complicated profiles, with $\gtrsim 
10$~components.

\section{Discussion}
\label{sec:discuss}

\subsection{Component statistics}
\label{sec:stat}

A total of 214 Gaussian components were detected in the 30 \hii\ absorption 
spectra, with the typical model consisting of several narrow components (which 
could be interpreted as arising in the CNM) and a few wide components 
(suggestive of absorption in warm gas). Note that, as argued in 
section~\ref{sec:cnm}, the spectra must contain absorption from gas with spin 
temperatures significantly higher than those typical of the CNM. Following 
\citet{heiles03a}, we will in this section refer to components with $\Tk \leq 
$~500\,K as arising in the CNM, to include possible effects of non-thermal 
broadening. Components with 500\,K~$\leq \Tk \leq 5000$\,K will be assumed to 
arise in the ``unstable'' phase, and components with $\Tk > 5000$\,K in the WNM.

We then find that $\sim 72$\% of the Gaussian components arise in the CNM, 
$\sim 25$\% in the ``unstable'' phase and $\sim 3\%$ in the WNM. We emphasize 
that we are merely counting components here, and not \hi\ column densities; 
this does {\it not} imply that 72\% of the \hi\ along our sightlines is in the 
CNM! If the analysis is restricted to sightlines with ``simple'' absorption 
profiles (defined here as having 7 or fewer Gaussian components, i.e. equal to 
or lower than the median number for the sample), there are a total of 93 
components, of which $\sim 69\%$, $\sim 26\%$ and $\sim 5\%$ are in the cold, 
unstable, and warm phases, respectively. Given that our spectra are 
sufficiently sensitive to detect gas in the classical WNM, it is somewhat 
surprising that we find so few components that are unambiguously in the warm 
phase. Interestingly, we also find that 14 (i.e. $\sim 7$\%) of the 
214~components have kinetic temperatures $\Tk < 40$\,K, i.e. below of the 
nominal temperature range of the the CNM phase. Such components have been 
detected in earlier studies \citep[e.g.][]{heiles03a,dickey03}. A simple 
explanation of their existence, in the context of two-phase models, is that 
photo-electric heating by the ejection of electrons from dust grains and 
polycyclic aromatic hydrocarbons is not important in these clouds, perhaps due 
to a lack of such large molecules \citep{wolfire95,heiles03b}. 

\subsection{Trends with Galactic latitude and longitude}
\label{sec:lb}

The absorption spectra of our sample cover a wide range of Galactic ($l,b$) 
values. The derived results from the sample are an average over all these 
sightlines, and are thus likely to sample a range of local conditions (e.g. 
density, pressure, background radiation field, etc). While the sample is as 
yet too small to test for variations of the derived parameters (e.g. the CNM 
fraction) with direction, we examine here whether the distribution of these 
parameters shows any striking trend with Galactic co-ordinates.

The distribution of peak optical depth, $\tau_{\rm peak}$, and velocity FWHM 
for the 214 absorption components is plotted versus Galactic latitude and 
longitude in the four panels of Fig.~\ref{fig:tvlb}. Larger peak optical depths 
are obtained at low Galactic latitudes (as expected), and also for sightlines 
near the Galactic anti-centre. The latter is interesting because the cold gas 
content is expected to decrease in the outskirts of the Galaxy, yielding lower 
peak optical depths. Of course, this is partly compensated for by the 
clustering in velocity space towards the anti-centre. Similarly, the typical 
number of components required to fit the profile increase as one approaches 
the Galactic plane. However, there is no significant correlation of the 
component widths with $l$ or $b$. This indicates that the widths of the 
individual absorption components do not have a significant contribution from 
Galactic rotation. 

\subsection{Non-thermal broadening}
\label{sec:non-therm}

\begin{figure}
\begin{center}
\includegraphics[scale=0.4]{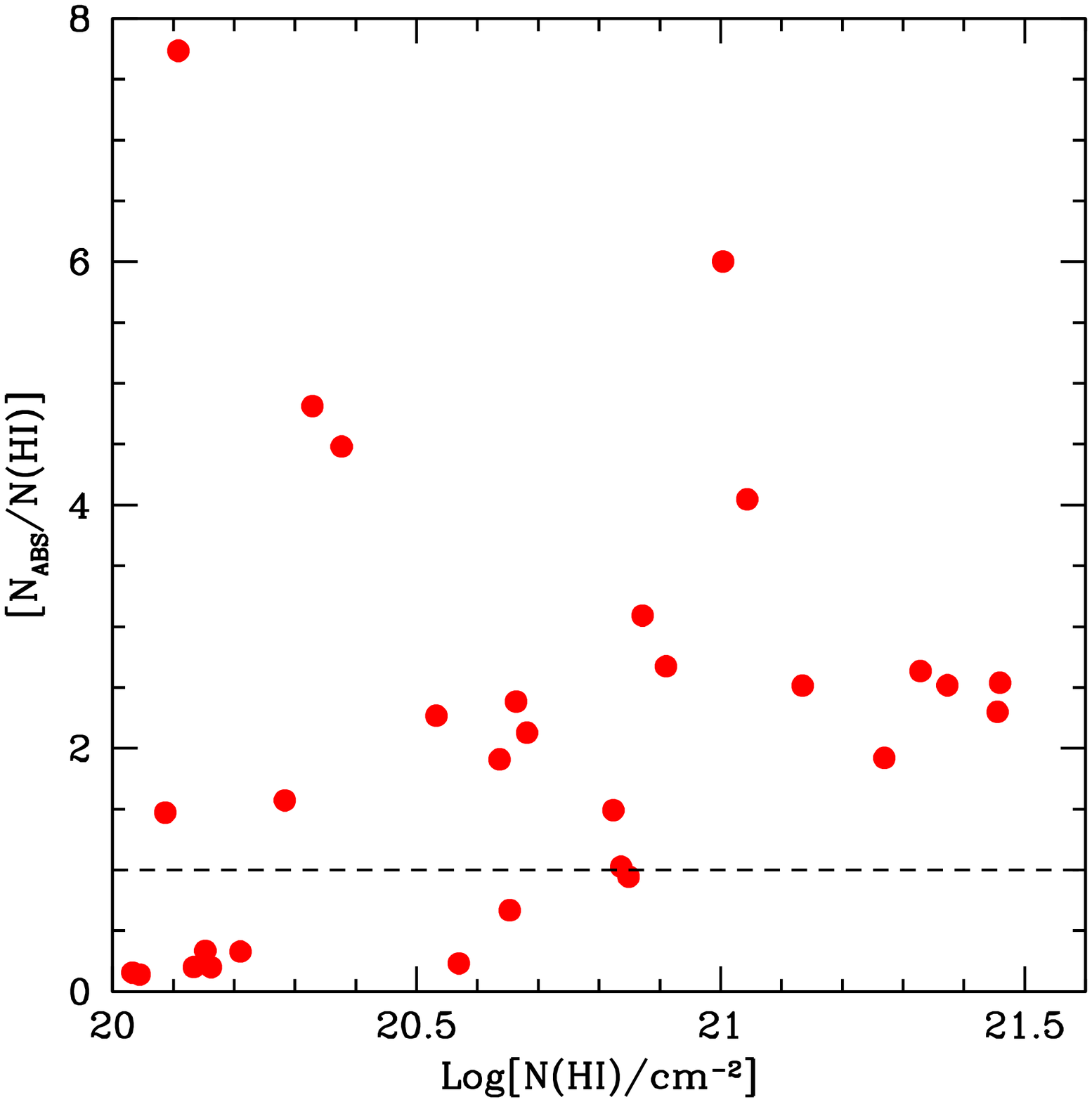}
\caption{\label{fig:non-therm} The figure shows the ratio N$_{\rm ABS}$/N(\hi) plotted against Log[N(\hi)] for each of the 30 sightlines, where N$_{\rm ABS}$ is the \hi\ column density derived from the fits to the absorption profile and N(\hi) is the \hi\ column density derived from the LAB emission profile, in the optically-thin limit. The ratio is typically $\approx 2$, except for a few low-N(\hi) sightlines. See discussion in Section~\ref{sec:non-therm}.}
\end{center}
\end{figure}

A test of the presence of non-thermal broadening within the Gaussian 
parametrisation can be carried out by comparing, for each sightline, the total 
\hi\ column density derived from the fit to the absorption spectrum N$_{\rm 
ABS}$ with the \hi\ column density N(\hi) measured from the LAB emission 
spectrum. The total \hi\ column density along a sightline from the Gaussian 
fit is given by the expression
\begin{equation}
{\rm N_{ABS}} = \Sigma_i 1.823 \times {\rm T}_{{\rm s},i} \times 1.06 \times \tau_{{\rm peak},i} \times \Delta V_i \:\:,
\label{eqn:nhi}
\end{equation}

\noindent where $\Delta V_i$ and $\tau_{{\rm peak},i}$ are the FWHM and the 
peak optical depth of the $i$'th component, respectively. Note that the 
Gaussian fit only yields the {\it doppler} temperature $\Td$, and not the {\it 
spin} temperature $\Ts$. Assuming $\Ts = \Td$ for each component yields an 
upper limit to its \hi\ column density, as $\Td \geq \Tk$ and $\Tk \geq \Ts$. 
Adding the contributions of all components along a sightline in 
equation~(\ref{eqn:nhi}) then gives an upper limit to the total \hi\ column 
density detected in absorption, N$_{\rm ABS}$, for the sightline. 

The ratio N$_{\rm ABS}$/N(\hi) is then an upper limit to the ratio of the \hi\ 
column density detected in \hii\ absorption to the total \hi\ column density. 
Of course, if only a small fraction of the total gas along a given sightline 
is detected in \hii\ absorption, N$_{\rm ABS}$ could be {\it lower} than 
N(\hi). Conversely, if non-thermal motions contribute to the component FWHMs, 
then the inferred doppler temperatures would be larger than the kinetic 
temperatures, which are, in turn, larger than or equal to the spin 
temperatures. This would imply that N$_{\rm ABS}$ would be {\it larger} than 
N(\hi). In other words, a scenario wherein the \hi\ column density derived 
from the absorption fits is systematically larger than the \hi\ column 
measured from the emission profile is indicative of non-thermal broadening 
(and/or that the spin temperature is lower than the kinetic temperature).

Fig.~\ref{fig:non-therm} plots the ratio $R \equiv$~N$_{\rm ABS}$/N(\hi) 
against N(\hi). Ten sightlines out of thirty have $R \lesssim 1$, as would be 
expected if non-thermal motions do not contribute appreciably to the component 
widths. However, fifteen sightlines have $R \approx 1.5-3$, while five 
sightlines have $R \approx 4-7.7$. We conclude that non-thermal broadening is 
likely to be present for twenty of the thirty sightlines. The median value of 
the ratio N$_{\rm ABS}$/N(\hi) is $\approx 2.0$. Note that this deviation from 
unity includes contributions both from non-thermal broadening and from the 
possibility that $\Ts < \Tk$. Thus, it appears that the contribution to the 
component widths from non-thermal broadening is larger than the thermal width 
by a factor $\lesssim \sqrt(3)$, and does not dominate the line widths.

Finally, one might infer the fraction of \hi\ in different phases of the ISM 
from the \hi\ column densities derived from the multi-Gaussian fits to the 
absorption profiles, as done by \citet{kanekar03b}. However, this requires one 
to use the doppler temperatures to differentiate between the CNM, WNM and 
thermally unstable neutral medium. Of course, the critical missing ingredient 
is the extent of non-thermal broadening of individual line components. For 
example, a significant fraction of absorption with doppler temperatures in the 
unstable range might arise from non-thermally broadened CNM absorption. We 
note that this is the approach followed by \citet{heiles03b}, based on their 
multi-Gaussian decomposition of the \hii\ absorption and emission profiles, to 
infer that around one-third of all \hi\ is in the thermally unstable phase. 
While \citet{heiles03b} identified all components that were not detected in 
absorption with the WNM, non-thermal broadening would reduce the strength of 
any absorption, possibly causing it to not be detected. It is hence difficult 
to disentangle the results of \citet{heiles03b} and \citet{kanekar03b} from 
the effects of non-thermal broadening. In the next section, we will attempt to 
more rigorously examine the presence or absence of unstable gas in the ISM.

\subsection{The two-phase model of the neutral ISM}
\label{sec:2-phase}

\begin{table*}
\begin{center}
\caption{Evidence for gas in the unstable temperature range. See text for discussion.}
\label{table:unstable}
 \begin{tabular}{lccccccc}
\hline
Source    &  N(\hi)            &  N$_{\rm CNM}$  & $f_{\rm CNM}^\ast$ & ${\rm T_{K,Max}}^\dagger$ & S/N$^\ddagger$ & $f_{\rm WNM}^{\ast\ast}$ & $f_{\rm UNM}^{\ast\ast\ast}$  \\
          & $\times 10^{20}$\,\cm\ & $\times 10^{20}$\,\cm &  &   K     &  &  & \\
          &       &  $\Ts = 200$\,K  &        &                  &  & &       \\
\hline
  B0023-263 &  1.62  &  0.093  & 0.06 & $ 1574 \pm  473$ &   3.0 &  0.95  &  0.00 \\
  B0114-211 &  1.42  &  0.494  & 0.35 & $  388 \pm  106$ &   1.5 &  1.00  &  0.00 \\
  B0117-155 &  1.45  &  0.112  & 0.08 & $  627 \pm  114$ &   3.0 &  0.92  &  0.00 \\
  B0134+329 &  4.33  &  1.615  & 0.37 & $ 3497 \pm  890$ &   WNM &  0.12  &  -    \\
  B0202+149 &  4.80  &  2.724  & 0.57 & $ 1347 \pm  475$ &   8.8 &  0.15  &  0.29 \\
  B0237-233 &  2.13  &  1.071  & 0.50 & $ 9468 \pm 1767$ &   WNM &  0.47  &  -    \\
  B0316+162 & 10.09  & 10.810  & 1.00 & $11364 \pm 1898$ &   WNM &  0.07  &  -   \\
  B0316+413 & 13.64  &  7.077  & 0.52 & $ 3683 \pm  196$ &  38.8 &  0.04  &  0.44 \\
  B0404+768 & 11.06  &  7.091  & 0.64 & $ 4718 \pm  182$ &   WNM &  0.17  &  -    \\
  B0407-658 &  3.41  &  1.998  & 0.59 & $ 4285 \pm 1576$ &   WNM &  0.43  &  -    \\
  B0518+165 & 23.60  & 22.750  & 0.96 & $ 2491 \pm  235$ &   2.4 &  0.04  &  0.00 \\
  B0531+194 & 28.80  & 14.810  & 0.51 & $ 3980 \pm  841$ &   WNM &  0.04  &  -    \\
  B0538+498 & 21.31  & 20.480  & 0.96 & $ 2693 \pm 2004$ &   WNM &  0.03  &  -    \\
  B0831+557 &  4.50  &  1.761  & 0.39 & $  664 \pm  124$ &   5.6 &  0.33  &  0.28 \\
  B0834-196 &  6.86  &  3.548  & 0.52 & $ 1045 \pm  186$ &   6.0 &  0.24  &  0.24 \\
  B0906+430 &  1.28  &  0.187  & 0.15 & $16320 \pm 4179$ &   WNM &  0.94  &  -    \\
  B1151-348 &  7.06  &  2.603  & 0.37 & $ 1104 \pm   91$ &   6.1 &  0.31  &  0.32 \\
  B1245-197 &  3.71  &  0.576  & 0.16 & $  529 \pm  249$ &   5.6 &  0.46  &  0.38 \\
  B1328+254 &  1.11  &  0.078  & 0.07 & $  559 \pm  149$ &   5.5 &  0.51  &  0.42 \\
  B1328+307 &  1.22  &  0.261  & 0.21 & $ 5143 \pm 1168$ &   WNM &  0.38  &  -    \\
  B1345+125 &  1.92  &  1.112  & 0.58 & $ 1691 \pm  220$ &   2.1 &  0.60  &  0.00 \\
  B1611+343 &  1.36  &  0.068  & 0.05 & $ 1037 \pm  278$ &   4.5 &  0.64  &  0.31 \\
  B1641+399 &  1.08  &  0.032  & 0.03 & $ 1350 \pm  492$ &   5.0 &  0.58  &  0.39 \\
  B1814-637 &  6.66  &  3.637  & 0.55 & $ 3433 \pm  388$ &   7.8 &  0.17  &  0.28 \\
  B1827-360 &  8.13  &  5.622  & 0.69 & $ 1678 \pm 1731$ &   WNM &  0.15  &  -    \\
  B1921-293 &  7.44  &  5.272  & 0.71 & $10605 \pm 3046$ &   WNM &  0.19  &  -    \\
  B2050+364 & 28.53  & 11.030  & 0.39 & $ 6246 \pm 5027$ &   WNM &  0.07  &  -    \\
  B2200+420 & 18.59  & 13.010  & 0.70 & $ 3632 \pm 3267$ &   WNM &  0.20  &  -    \\
  B2203-188 &  2.38  &  0.905  & 0.38 & $14617 \pm 5183$ &   WNM &  0.43  &  -    \\
  B2223-052 &  4.61  &  3.770  & 0.82 & $ 1405 \pm  105$ &   3.6 &  0.15  &  0.03 \\
\hline
\end{tabular}
\end{center}
\begin{flushleft}
$\dagger$~The kinetic temperature of the widest fitted component along each sightline. In the case of B0538$+$498 and B1827$-$360, the listed component has a lower doppler temperature than that of another component along the sightline (which have $\Tk = 3552 \pm 342$\,K and $\Tk = 1699 \pm 165$~K, respectively). However, the $1\sigma$ error is larger for the listed components, due to which their kinetic temperatures are formally consistent with the WNM range. We have hence chosen to list these components in the table.\\
$^\ddagger$~The signal-to-noise ratio at which residual \hi\ (i.e. after subtracting out N$_{\rm CNM}$) would have been detected in our \hii\ absorption spectra, for an assumed $\Ts = 5000$\,K and line FWHM~$=\Delta V_{90}$. Systems with detected WNM components are listed as ``WNM'' in this column. \\
$^{\ast}$~The upper limit to the fraction of gas in the CNM phase along the sightline, assuming that all the absorption arises in CNM with a spin temperature of $200$\,K.\\
$^{\ast\ast}$~The upper limit to the fraction of gas in the stable WNM phase along the sightline, for a $3\sigma$ detection significance, assuming $\Ts = 5000$~K and line FWHM~$= \Delta V_{90}$.\\
$^{\ast\ast\ast}$~The lower limit to the fraction of gas in the unstable phase ($200 < \Tk < 5000$\,K), assuming that all detected \hii\ absorption arises in the CNM and that there is stable WNM along the sightline just below the $3\sigma$ detection threshold for absorption. This is essentially $1 - (f_{\rm CNM} + f_{\rm WNM})$. For sources with detected WNM components, this column contains a ``-''.  \\
\end{flushleft}
\end{table*}

\begin{figure}
\begin{center}
\includegraphics[scale=0.4]{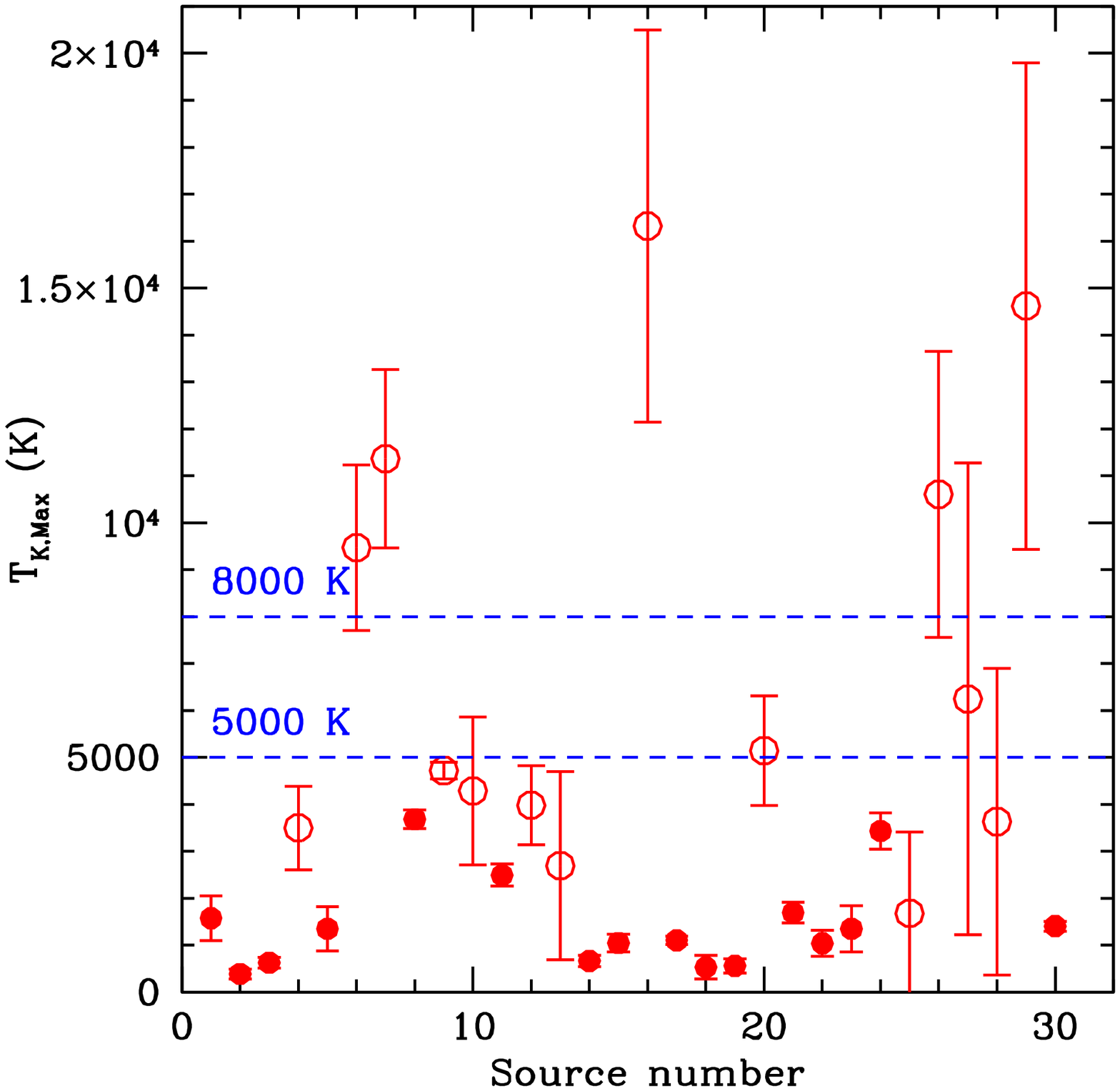}
\caption{\label{fig:wnmmax} The maximum kinetic temperature of the different components on each of the 30 sightlines, from the multi-Gaussian fits; the stable WNM temperature range ($5000 - 8000$\,K) is indicated by the horizontal dashed lines. Components with temperatures consistent with stable WNM are indicated by open circles, while those definitely outside the above range (at $\geq 3\sigma$ significance) are shown as filled circles. See Section~\ref{sec:2-phase} for discussion.}
\end{center}
\end{figure}

In Section~\ref{sec:cnm} we showed that the detected \hii\ absorption cannot 
arise solely in gas with spin temperatures in the CNM range, but must include 
absorption from warmer gas. In other words, it is reasonable to associate wide 
components that are found in the Gaussian decomposition of our spectra with 
warm gas, as opposed to non-thermally broadened CNM. In the classic two-phase 
models, this additional absorption can only arise in stable WNM. However, only 
$\approx 5$\% of the Gaussian components have kinetic temperatures consistent 
with values expected in the WNM. Fig.~\ref{fig:wnmmax} compares the maximum 
kinetic temperature obtained along each sightline (${\rm T_{k,Max}}$) to the 
stable WNM temperature range. It is clear that only five sightlines have 
maximum inferred kinetic temperatures consistent with absorption by stable WNM 
($5000 \leq \Tk \leq 8000$~K), within $1\sigma$ significance. An additional 
nine sightlines have maximum inferred kinetic temperatures consistent with the 
above temperature range within $3\sigma$ significance. Even if we include all 
of these as possible WNM components, there are sixteen sightlines where the 
highest doppler temperature is significantly lower than the stable range of 
WNM temperatures. 

For the sixteen sightlines with no detected absorption component with $\Td$ in 
the stable WNM range, it is possible that all the observed \hii\ absorption 
arises in non-thermally broadened CNM. Assuming that such non-thermally 
broadened CNM has a spin temperature of 200~K yields an upper limit to the CNM 
column density along each sightline. If the \hi\ indeed exists in two stable 
phases, the CNM and the WNM, with no gas in the thermally unstable phase, then 
the rest of the \hi\ detected in the \hi\ emission spectrum on each sightline 
must be stable WNM. Following the arguments in Section~\ref{sec:cnm}, one can 
immediately test whether this residual \hi\ would have been detected in the 
\hii\ absorption spectrum. For this purpose, we again assume, conservatively, 
that the WNM has $\Ts = 5000$~K and a line FWHM~$= \Delta V_{90}$. 

These results are summarized in Table~\ref{table:unstable} whose columns are 
(1)~the source name, (2)~the total \hi\ column density, from the isothermal 
estimate N(\hi), (3)~the upper limit to the CNM column density, ${\rm N_{\rm 
CNM}}$, for $\Ts = 200$\,K, (4)~the upper limit to the CNM fraction along the 
sightline, using the ratio ${\rm N_{CNM}}/$N(\hi), (5)~the maximum kinetic 
temperature measured along the sightline, (6)~for the sixteen systems with no 
absorption components corresponding to stable WNM, the S/N at which the 
residual \hi\ would have been detected, for line FWHM=$\Delta V_{90}$ and $\Ts 
= 5000$\,K, (7)~for these sightlines, the upper limit to the fraction of \hi\ 
in the stable WNM (assuming a $3\sigma$ detection significance), and (8)~for 
these sightlines, the lower limit to the fraction of \hi\ in the unstable 
phase, assuming that all the \hii\ absorption arises in the CNM and that the 
WNM column is just below the $3\sigma$ detection threshold for \hii\ 
absorption. Sources with detected WNM components are listed as ``WNM'' in the 
sixth column and ``-'' in the last column of the table.

We emphasize that the assumptions are extremely conservative, using (1)~the 
highest spin temperature in the CNM range, (2)~the highest spin temperature 
expected in the WNM temperature range \citep{liszt01}, and (3)~assuming 
extreme non-thermal broadening of the putative WNM components. All of these 
would reduce the strength (and hence the detection S/N) of the expected WNM 
absorption. Even with these assumptions, we find that \hii\ absorption would 
have been detected at $\geq 3\sigma$ significance on thirteen of the sixteen 
sightlines if the residual gas along the sightline was in the stable WNM phase 
(and at $\geq 4.5\sigma$ significance on ten of the sightlines). In other 
words, for these thirteen sightlines, if all the residual gas is WNM, the 
multi-Gaussian fit would have included at least one component with doppler 
temperature at or above the stable WNM range. Thus, the spectra of thirteen 
(i.e. $\sim 43$\%) out of the thirty sightlines are inconsistent with a model 
in which neutral gas exists in only the stable temperature ranges of the 
two-phase models. We conclude that, along these thirteen sightlines, much of 
the difference between the total \hi\ column density and the upper limit to 
the CNM column density must arise in gas in the thermally unstable phase.

The non-detection of WNM components along the above thirteen sightlines can be 
used to place an upper limit on the WNM column density along these sightlines, 
and hence, to obtain a $3\sigma$ upper limit on the WNM fraction $f_{\rm WNM}$. 
We also already have an upper limit to the CNM column density and the CNM 
fraction along these sightlines by assuming that all the detected \hii\ 
absorption arises in cold gas with $\Ts = 200$~K. The difference between the 
total \hi\ column density and the sum of these {\it upper limits} on the WNM 
and CNM column densities then yields a {\it lower limit} to the column density 
of gas in the unstable neutral medium, and thus, a lower limit to the fraction 
$f_{\rm UNM}$ of the gas in the unstable phase (at $200$\,K~$ < \Tk < 
5000$\,K). The latter lower limits are listed in the last column of 
Table~\ref{table:unstable}. For the above thirteen sightlines where stable WNM 
would, if present, have been detected at $\geq 3\sigma$ significance, the 
lower limits to the fraction of unstable gas lie in the range $0.00 < f_{\rm 
UNM} < 0.44$, with a median value of $\approx 0.28$. We emphasize that these 
are {\it lower limits} to $f_{\rm UNM}$. The assumption that all the observed 
\hii\ absorption arises in the CNM is obviously incorrect, since \hii\ 
absorption from gas in the unstable phase would have lower spin temperatures 
and velocity widths than absorption by stable WNM, and would hence have been 
detected at higher significance. Indeed, each of the thirteen sightlines 
contains at least one component with doppler temperature in the range 
$200$~K~$< \Td < 5000$~K. Thus, some of the observed \hii\ absorption must 
arise in gas in the unstable temperature range. In addition, the high assumed 
CNM spin temperature implies that the listed CNM fractions are all stringent 
upper limits. Finally, the WNM fractions are also stringent upper limits 
because we have assumed that the WNM column density is just below our 
$3\sigma$ detection threshold, for $\Ts = 5000$~K and line FWHM~$= \Delta 
V_{90}$. {\it We hence conclude that a significant fraction of the gas, 
typically $\gtrsim 28$\%, must be present in the unstable phase along thirteen 
of our sightlines, and that the ``classic'' two-phase model is not viable for 
at least these sightlines.} We also note, in passing, that it is likely that 
many of the remaining seventeen sightlines also contain thermally unstable 
gas, given that almost all of them contain absorption with $\Td$ in the range 
$500 < \Td < 5000$~K.

For comparison, \citet{heiles03b} found $> 30$\% of the \hi\ along their 
sightlines to have temperatures in the unstable range (which they defined as 
$500$\,K~$< \Td < 5000$\,K), in reasonable agreement with our results. This is 
curious, given the fact that their derived fractions in different phases would 
be seriously affected by non-thermal broadening, which was not accounted for 
in their analysis. Note that our analysis explicitly takes into account the 
possibility of contributions from non-thermal broadening to the line widths, 
via the conservative assumption that the WNM line widths are equal to $\Delta 
V_{90}$, the width of the \hii\ emission profile that contributes 90\% of the 
\hii\ emission.

\section{Conclusions}
\label{sec:conclude}

We have studied the temperature distribution of the neutral interstellar 
medium along 33 Galactic sightlines, using deep, high velocity resolution 
($\approx 0.26 - 0.52$~\kms) \hii\ absorption spectra obtained with the WSRT, 
the GMRT and the ATCA. The typical RMS optical depth noise of the spectra is 
$\lesssim 10^{-3}$ per 1~\kms\ velocity channel, making the spectra 
sufficiently sensitive to detect \hii\ absorption from gas in the warm neutral 
medium, with \hi\ column densities $\gtrsim 10^{20}$cm$^{-2}$ and velocity 
widths comparable to the thermal width. 

We used conservative assumptions regarding the CNM spin temperature ($\Ts = 
200$~K), the WNM spin temperature ($\Ts = 5000$~K) and non-thermal broadening 
of the WNM (line FWHM~$= \Delta V_{90}$) to consider the canonical two-phase 
model of the ISM wherein the \hii\ absorption features arise in the CNM, while 
the emission profiles contain contributions from both stable CNM and WNM. We 
find that such a model can be ruled out for more than half our sightlines. 
Some of the \hii\ absorption that we detect {\it must} arise from gas with 
spin temperatures larger than that of the CNM (i.e. must arise in the WNM or 
with gas with temperatures in the thermally unstable range). 

We then used a multi-Gaussian decomposition of 30 of the \hii\ absorption 
spectra to examine the possibility that some Galactic \hi\ might be in the 
unstable temperature range, $200 < \Tk < 5000$~K. We found very few components 
with doppler line widths consistent with values expected from stable or 
non-thermally broadened WNM. For sixteen sightlines, we found no \hii\ 
absorption components with line widths consistent with an origin in the WNM. 
Thirteen of the sixteen spectra are sufficiently sensitive to detect any WNM 
along the sightline even if extreme non-thermal broadening results in line 
FWHMs comparable to the $\Delta V_{90}$ of the \hii\ emission profile. We can 
thus rule out the possibility that the neutral hydrogen along these sixteen 
sightlines is in two phases, with temperatures in the stable CNM and WNM 
ranges. Some fraction of the \hii\ absorption that we detect on these 
sightlines must arise from gas with temperatures in the intermediate, 
thermally unstable range (at $200$\,K~$ < \Tk < 5000$\,K) of the two-phase 
model. For the above thirteen sightlines, we obtain a median lower limit of 
$28$\% to the gas fraction in this unstable phase. Our observations hence 
robustly indicate that a significant fraction of the gas in the Galactic ISM 
has a temperature outside the thermally stable ranges in two-phase models.

\section*{Acknowledgements}

This research has made use of the NASA's Astrophysics Data System. We thank 
Robert Braun for useful discussions and his comments on an earlier version of 
the paper. We are very grateful to Harvey Liszt for providing us his simulation 
results, and to Ranjeev Misra, Rajaram Nityananda, Sanjay Bhatnagar and A. 
Pramesh Rao for many helpful comments. We thank the staff of the GMRT and the 
WSRT who have made these observations possible. The GMRT is run by the 
National Centre for Radio Astrophysics of the Tata Institute of Fundamental 
Research. The WSRT is operated by ASTRON (the Netherlands Institute for Radio 
Astronomy), with support from the Netherlands Foundation for Scientific 
Research (NWO). Some of the data used in this paper were obtained from the 
Leiden/Argentine/Bonn Galactic \hi\ Survey. NR acknowledges the Jansky 
Fellowship Program of NRAO/NSF/AUI and support from the Alexander von Humboldt 
Foundation. NR also acknowledges support from NCRA during his Ph.D., when a 
significant fraction of this work was done. NK acknowledges support from the 
Department of Science and Technology through a Ramanujan Fellowship.

\bibliographystyle{mn2e}

\bsp

\appendix

\section{The results of the Gaussian fits}
\label{sec:app1}

The parameters of the multi-Gaussian fits are summarised in Table~A1. The four 
columns of the table list (1)~the peak \hii\ optical depth $\tau_{\rm peak}$, 
(2)~the central absorption velocity $V_c$, in \kms, (3)~the $b$-parameter, in 
\kms, and (4)~the doppler temperature $\Td$, in K.

\setcounter{table}{0}
\begin{table}
\caption{The parameters of the multi-Gaussian fits.}
\label{table:fit}
\vskip -0.15in
\begin{center}
 \begin{tabular}{cr@{ $\pm$}rcr@{ $\pm$}r}
 \hline
$\tau_{\rm peak}$ & \multicolumn{2}{c}{$V_c$} & $b$ & \multicolumn{2}{c}{$\Td$} \\
 & \multicolumn{2}{c}{km~s$^{-1}$} & km~s$^{-1}$ & \multicolumn{2}{c}{K}\\
\hline\\[-1.5ex]
\multicolumn{6}{c}{B0023$-$263 (WSRT); ~~$N_{\rm comp} = ~1$; ~~$\chi_r^2=0.999$:}\\
\hline
  0.00207 $\pm$  0.00027 &     -5.09 &      0.54 &    5.10 $\pm$    0.77 &   1574 &    473 \\
\hline\\[-1.5ex]
\multicolumn{6}{c}{B0114$-$211 (WSRT); ~~$N_{\rm comp} = ~3$; ~~$\chi_r^2=1.004$:}\\
\hline
   0.0327 $\pm$   0.0037 &    -7.727 &     0.038 &   1.109 $\pm$   0.086 &     75 &     12 \\
  0.00620 $\pm$  0.00089 &     -2.07 &      0.14 &    1.08 $\pm$    0.20 &     70 &     26 \\
   0.0119 $\pm$   0.0038 &     -7.10 &      0.25 &    2.53 $\pm$    0.35 &    388 &    106 \\
\hline\\[-1.5ex]
\multicolumn{6}{c}{B0117$-$155 (WSRT); ~~$N_{\rm comp} = ~1$; ~~$\chi_r^2=1.007$:}\\
\hline
  0.00448 $\pm$  0.00035 &     -6.33 &      0.21 &    3.22 $\pm$    0.29 &    627 &    114 \\
\hline\\[-1.5ex]
\multicolumn{6}{c}{B0134$+$329 (WSRT); ~~$N_{\rm comp} = ~9$; ~~$\chi_r^2=1.055$:}\\
\hline
  0.00065 $\pm$  0.00024 &     19.99 &      0.45 &    1.49 $\pm$    0.64 &    135 &    119 \\
   0.0518 $\pm$   0.0011 &     2.258 &     0.013 &   1.489 $\pm$   0.025 &  134.4 &    4.7 \\
   0.0131 $\pm$   0.0014 &    -0.663 &     0.050 &   1.072 $\pm$   0.094 &     70 &     13 \\
   0.0109 $\pm$   0.0013 &    -2.413 &     0.051 &   0.774 $\pm$   0.083 &   36.3 &    8.0 \\
   0.0077 $\pm$   0.0019 &     -2.83 &      0.67 &    7.60 $\pm$    0.94 &   3497 &    890 \\
   0.0140 $\pm$   0.0017 &     -4.56 &      0.11 &    2.23 $\pm$    0.23 &    301 &     63 \\
  0.01651 $\pm$  0.00074 &   -11.104 &     0.038 &   1.685 $\pm$   0.070 &    172 &     15 \\
  0.00756 $\pm$  0.00040 &    -15.33 &      0.20 &    3.79 $\pm$    0.23 &    870 &    107 \\
  0.00093 $\pm$  0.00017 &    -29.27 &      0.48 &    3.29 $\pm$    0.68 &    656 &    278 \\
\hline\\[-1.5ex]
\multicolumn{6}{c}{B0202$+$149 (WSRT); ~~$N_{\rm comp} = ~7$; ~~$\chi_r^2=0.923$:}\\
\hline
   0.0570 $\pm$   0.0045 &   -10.761 &     0.042 &   1.434 $\pm$   0.052 &  124.7 &    8.7 \\
   0.0428 $\pm$   0.0059 &    -5.749 &     0.039 &    1.63 $\pm$    0.10 &    160 &     19 \\
   0.0164 $\pm$   0.0019 &    -0.120 &     0.071 &    1.74 $\pm$    0.14 &    183 &     28 \\
   0.0129 $\pm$   0.0060 &    -13.53 &      0.37 &    2.05 $\pm$    0.47 &    256 &    112 \\
  0.00189 $\pm$  0.00042 &    -52.76 &      0.31 &    1.72 $\pm$    0.44 &    180 &     88 \\
   0.0446 $\pm$   0.0050 &     -7.09 &      0.54 &    4.71 $\pm$    0.86 &   1347 &    475 \\
  0.00230 $\pm$  0.00054 &      3.79 &      0.32 &    1.44 $\pm$    0.45 &    126 &     76 \\
\hline\\[-1.5ex]
\multicolumn{6}{c}{B0237$-$233 (WSRT); ~~$N_{\rm comp} = ~3$; ~~$\chi_r^2=0.963$:}\\
\hline
  0.11190 $\pm$  0.00072 &   -0.0689 &    0.0051 &  0.9871 $\pm$  0.0078 &  59.07 &   0.91 \\
  0.00836 $\pm$  0.00051 &    -10.01 &      0.13 &    2.78 $\pm$    0.21 &    468 &     71 \\
  0.00255 $\pm$  0.00034 &      -4.5 &       1.1 &    12.5 $\pm$     1.2 &   9468 &   1767 \\
\hline\\[-1.5ex]
\multicolumn{6}{c}{B0316$+$162 (GMRT); ~~$N_{\rm comp} = ~8$; ~~$\chi_r^2=1.190$:}\\
\hline
   0.2614 $\pm$   0.0061 &    -2.666 &     0.050 &   1.660 $\pm$   0.034 &  167.0 &    7.6 \\
   0.0627 $\pm$   0.0028 &     6.007 &     0.065 &   2.518 $\pm$   0.085 &    384 &     28 \\
   0.0313 $\pm$   0.0018 &     7.025 &     0.023 &   0.913 $\pm$   0.044 &   50.5 &    5.4 \\
  0.01829 $\pm$  0.00056 &   -10.188 &     0.026 &   1.096 $\pm$   0.042 &   72.8 &    6.1 \\
   0.0066 $\pm$   0.0015 &     -1.13 &      0.55 &    13.7 $\pm$     1.0 &  11364 &   1898 \\
   0.0540 $\pm$   0.0028 &      0.74 &      0.23 &    5.83 $\pm$    0.25 &   2058 &    192 \\
   0.0780 $\pm$   0.0095 &    -1.731 &     0.019 &   0.747 $\pm$   0.043 &   33.8 &    4.2 \\
   0.4122 $\pm$   0.0050 &    -0.125 &     0.013 &   1.383 $\pm$   0.013 &  116.0 &    2.4 \\
\hline\\[-1.5ex]
\multicolumn{6}{c}{B0316$+$413 (WSRT); ~~$N_{\rm comp} = 14$; ~~$\chi_r^2=0.988$:}\\
\hline
  0.00109 $\pm$  0.00029 &    -34.51 &      0.44 &    2.15 $\pm$    0.76 &    280 &    198 \\
  0.00047 $\pm$  0.00016 &    -55.78 &      0.72 &     2.6 $\pm$     1.0 &    411 &    318 \\
  0.00055 $\pm$  0.00019 &    -39.78 &      0.67 &     2.1 $\pm$     1.0 &    255 &    250 \\
   0.0235 $\pm$   0.0091 &     3.378 &     0.094 &    1.07 $\pm$    0.15 &     69 &     20 \\
   0.0492 $\pm$   0.0037 &    -2.155 &     0.065 &   1.094 $\pm$   0.092 &     73 &     12 \\
   0.0283 $\pm$   0.0032 &      0.08 &      0.16 &    7.79 $\pm$    0.21 &   3683 &    196 \\
  0.01769 $\pm$  0.00067 &     8.978 &     0.046 &   1.871 $\pm$   0.068 &    212 &     15 \\
\hline
\end{tabular}\\
\end{center}
\end{table}
\setcounter{table}{0}
\begin{table}
\caption{({\it continued}) The parameters of the multi-Gaussian fits.}
\vskip -0.15in
\begin{center}
 \begin{tabular}{cr@{ $\pm$}rcr@{ $\pm$}r}
\hline
$\tau_{\rm peak}$ & \multicolumn{2}{c}{$V_c$} & $b$ & \multicolumn{2}{c}{$\Td$} \\
 & \multicolumn{2}{c}{km~s$^{-1}$} & km~s$^{-1}$ & \multicolumn{2}{c}{K}\\
\hline\\[-1.5ex]
\multicolumn{6}{c}{B0316$+$413 (WSRT); ~~$N_{\rm comp} = 14$; ~~$\chi_r^2=0.988$ (cont.):}\\
\hline
   0.1952 $\pm$   0.0068 &      2.20 &      0.10 &   2.222 $\pm$   0.084 &    299 &     22 \\
   0.1139 $\pm$   0.0098 &    -0.380 &     0.024 &   1.107 $\pm$   0.060 &   74.2 &    8.1 \\
   0.0168 $\pm$   0.0018 &    -4.811 &     0.027 &   0.689 $\pm$   0.054 &   28.8 &    4.5 \\
   0.0146 $\pm$   0.0014 &   -21.794 &     0.032 &    2.10 $\pm$    0.12 &    268 &     29 \\
  0.00535 $\pm$  0.00026 &    -26.11 &      0.82 &     5.3 $\pm$     1.1 &   1697 &    687 \\
   0.0528 $\pm$   0.0020 &    -5.302 &     0.072 &   2.276 $\pm$   0.059 &    314 &     16 \\
  0.00313 $\pm$  0.00027 &    -15.30 &      0.28 &    2.96 $\pm$    0.38 &    530 &    136 \\
\hline\\[-1.5ex]
\multicolumn{6}{c}{B0404$+$768 (WSRT); ~~$N_{\rm comp} = ~8$; ~~$\chi_r^2=0.988$:}\\
\hline
  0.00261 $\pm$  0.00053 &     11.57 &      0.26 &    1.54 $\pm$    0.38 &    144 &     70 \\
  0.06115 $\pm$  0.00092 &     3.183 &     0.029 &   1.566 $\pm$   0.040 &  148.6 &    7.5 \\
   0.2533 $\pm$   0.0081 &   -0.4963 &    0.0082 &   0.959 $\pm$   0.017 &   55.8 &    2.0 \\
  0.08553 $\pm$  0.00097 &   -10.897 &     0.017 &   1.328 $\pm$   0.027 &  106.9 &    4.3 \\
  0.04637 $\pm$  0.00082 &   -13.606 &     0.031 &   1.326 $\pm$   0.040 &  106.6 &    6.4 \\
   0.1497 $\pm$   0.0076 &    -0.901 &     0.035 &   1.934 $\pm$   0.045 &    227 &     10 \\
   0.0156 $\pm$   0.0013 &    -6.490 &     0.046 &    1.03 $\pm$    0.11 &     64 &     14 \\
   0.0307 $\pm$   0.0014 &     -4.49 &      0.17 &    8.82 $\pm$    0.17 &   4718 &    182 \\
\hline\\[-1.5ex]
\multicolumn{6}{c}{B0407$-$658 (ATCA); ~~$N_{\rm comp} = ~7$; ~~$\chi_r^2=0.981$:}\\
\hline
   0.1117 $\pm$   0.0027 &   16.3923 &    0.0089 &   1.047 $\pm$   0.018 &   66.4 &    2.3 \\
  0.00228 $\pm$  0.00057 &      19.8 &       1.4 &     8.4 $\pm$     1.6 &   4285 &   1576 \\
  0.02588 $\pm$  0.00077 &     0.497 &     0.029 &   1.304 $\pm$   0.049 &  103.1 &    7.6 \\
  0.00144 $\pm$  0.00025 &      49.1 &       1.1 &     7.8 $\pm$     1.6 &   3667 &   1460 \\
  0.00280 $\pm$  0.00046 &     -0.52 &      0.75 &     7.5 $\pm$     1.2 &   3437 &   1124 \\
   0.0417 $\pm$   0.0026 &    17.053 &     0.058 &   2.348 $\pm$   0.076 &    334 &     21 \\
  0.00265 $\pm$  0.00057 &    -18.26 &      0.27 &    1.52 $\pm$    0.38 &    140 &     69 \\
\hline\\[-1.5ex]
\multicolumn{6}{c}{B0518$+$165 (WSRT); ~~$N_{\rm comp} = 13$; ~~$\chi_r^2=1.048$:}\\
\hline
  0.02574 $\pm$  0.00076 &   -22.138 &     0.021 &   1.039 $\pm$   0.038 &   65.4 &    4.9 \\
  0.01885 $\pm$  0.00054 &   -21.585 &     0.066 &    5.21 $\pm$    0.12 &   1645 &     79 \\
  0.00379 $\pm$  0.00059 &    -12.89 &      0.15 &    1.15 $\pm$    0.23 &     80 &     33 \\
    0.221 $\pm$    0.021 &    -0.675 &     0.030 &   0.945 $\pm$   0.048 &   54.1 &    5.6 \\
   0.0266 $\pm$   0.0020 &     -6.12 &      0.24 &    3.93 $\pm$    0.20 &    936 &    100 \\
    0.136 $\pm$    0.013 &     -2.01 &      0.12 &   1.242 $\pm$   0.080 &     93 &     12 \\
   0.2653 $\pm$   0.0057 &     1.051 &     0.027 &   1.151 $\pm$   0.029 &   80.3 &    4.2 \\
   0.0841 $\pm$   0.0027 &      4.85 &      0.34 &    6.41 $\pm$    0.29 &   2491 &    235 \\
   1.0378 $\pm$   0.0073 &     5.968 &     0.012 &   1.162 $\pm$   0.011 &   81.8 &    1.6 \\
    0.327 $\pm$    0.017 &     7.354 &     0.018 &   0.762 $\pm$   0.024 &   35.2 &    2.2 \\
   0.4060 $\pm$   0.0042 &     8.965 &     0.018 &   1.488 $\pm$   0.019 &  134.2 &    3.6 \\
  0.00102 $\pm$  0.00033 &     21.91 &      0.83 &     3.0 $\pm$     1.3 &    532 &    490 \\
  0.00107 $\pm$  0.00042 &     28.13 &      0.56 &    1.63 $\pm$    0.81 &    160 &    163 \\
\hline\\[-1.5ex]
\multicolumn{6}{c}{B0531$+$194 (GMRT); ~~$N_{\rm comp} = 11$; ~~$\chi_r^2=1.123$:}\\
\hline
   0.0021 $\pm$   0.0011 &     -10.3 &       1.9 &     4.5 $\pm$     1.7 &   1200 &    972 \\
   0.0111 $\pm$   0.0030 &     -2.43 &      0.17 &    1.46 $\pm$    0.27 &    129 &     50 \\
  0.00199 $\pm$  0.00051 &     -27.6 &       1.8 &     3.4 $\pm$     1.8 &    693 &    767 \\
   0.0084 $\pm$   0.0012 &    -23.13 &      0.27 &    2.39 $\pm$    0.25 &    345 &     76 \\
    0.387 $\pm$    0.014 &    1.4574 &    0.0085 &   1.142 $\pm$   0.018 &   79.0 &    2.7 \\
    0.229 $\pm$    0.020 &      2.55 &      0.16 &    2.69 $\pm$    0.17 &    440 &     57 \\
    0.057 $\pm$    0.017 &      4.29 &      0.34 &    8.10 $\pm$    0.81 &   3980 &    841 \\
    0.151 $\pm$    0.017 &     6.202 &     0.054 &    1.62 $\pm$    0.11 &    159 &     22 \\
   0.2714 $\pm$   0.0092 &     9.213 &     0.016 &   1.537 $\pm$   0.031 &  143.1 &    6.2 \\
   0.0301 $\pm$   0.0040 &    12.050 &     0.099 &    1.38 $\pm$    0.13 &    116 &     23 \\
  0.00624 $\pm$  0.00048 &     21.05 &      0.13 &    2.06 $\pm$    0.21 &    256 &     55 \\
\hline
\end{tabular}\\
\end{center}
\label{table:fit1}
\end{table}
\setcounter{table}{0}
\begin{table}
\caption{({\it continued}) The parameters of the multi-Gaussian fits.}
\vskip -0.15in
\begin{center}
 \begin{tabular}{cr@{ $\pm$}rcr@{ $\pm$}r}
\hline
$\tau_{\rm peak}$ & \multicolumn{2}{c}{$V_c$} & $b$ & \multicolumn{2}{c}{$\Td$} \\
 & \multicolumn{2}{c}{km~s$^{-1}$} & km~s$^{-1}$ & \multicolumn{2}{c}{K}\\
\hline\\[-1.5ex]
\multicolumn{6}{c}{B0538$+$498 (WSRT); ~~$N_{\rm comp} = 20$; ~~$\chi_r^2=1.095$:}\\
\hline
   0.0645 $\pm$   0.0024 &     4.720 &     0.022 &   1.251 $\pm$   0.023 &   94.9 &    3.7 \\
    0.336 $\pm$    0.010 &      0.79 &      0.12 &    2.44 $\pm$    0.10 &    360 &     32 \\
    0.211 $\pm$    0.016 &   -0.4550 &    0.0094 &   1.087 $\pm$   0.024 &   71.6 &    3.3 \\
   0.0278 $\pm$   0.0024 &     -1.94 &      0.54 &    7.66 $\pm$    0.37 &   3554 &    359 \\
    0.084 $\pm$    0.018 &     -2.32 &      0.13 &   1.730 $\pm$   0.083 &    181 &     18 \\
    0.718 $\pm$    0.080 &    -8.091 &     0.042 &   0.869 $\pm$   0.015 &   45.8 &    1.6 \\
    0.210 $\pm$    0.065 &     -8.92 &      0.19 &   0.917 $\pm$   0.086 &     51 &     10 \\
   0.1561 $\pm$   0.0059 &   -11.022 &     0.010 &   0.989 $\pm$   0.026 &   59.3 &    3.3 \\
   0.1584 $\pm$   0.0066 &   -11.267 &     0.080 &   3.743 $\pm$   0.061 &    849 &     29 \\
   0.0449 $\pm$   0.0028 &   -13.855 &     0.030 &   1.089 $\pm$   0.043 &   71.9 &    5.9 \\
  0.01407 $\pm$  0.00025 &   -19.664 &     0.019 &   1.367 $\pm$   0.031 &  113.3 &    5.4 \\
  0.00232 $\pm$  0.00013 &    -27.42 &      0.43 &    5.89 $\pm$    0.78 &   2103 &    584 \\
  0.00236 $\pm$  0.00030 &    -34.51 &      0.13 &    1.58 $\pm$    0.23 &    152 &     47 \\
  0.00216 $\pm$  0.00018 &    -41.16 &      0.71 &     6.6 $\pm$     1.4 &   2643 &   1136 \\
  0.00273 $\pm$  0.00028 &   -43.591 &     0.087 &    1.21 $\pm$    0.15 &     89 &     24 \\
  0.00077 $\pm$  0.00030 &    -49.44 &      0.21 &    0.66 $\pm$    0.32 &     27 &     27 \\
 0.000644 $\pm$ 0.000095 &     -55.8 &       1.5 &     6.7 $\pm$     2.4 &   2693 &   2004 \\
 0.000337 $\pm$ 0.000090 &     -73.0 &       1.5 &     6.3 $\pm$     2.3 &   2378 &   1841 \\
  0.00065 $\pm$  0.00012 &   -115.83 &      0.51 &    3.29 $\pm$    0.73 &    656 &    306 \\
  0.00373 $\pm$  0.00016 &  -124.236 &     0.068 &   1.941 $\pm$   0.097 &    228 &     24 \\
\hline\\[-1.5ex]
\multicolumn{6}{c}{B0831$+$557 (WSRT); ~~$N_{\rm comp} = ~7$; ~~$\chi_r^2=1.000$:}\\
\hline
  0.00194 $\pm$  0.00047 &     -5.59 &      0.63 &    2.25 $\pm$    0.84 &    308 &    229 \\
   0.0596 $\pm$   0.0065 &     3.355 &     0.035 &   1.238 $\pm$   0.063 &   92.9 &    9.4 \\
   0.0289 $\pm$   0.0072 &     0.959 &     0.074 &    1.32 $\pm$    0.13 &    105 &     21 \\
   0.0321 $\pm$   0.0094 &      1.94 &      0.11 &    3.31 $\pm$    0.31 &    664 &    124 \\
  0.02651 $\pm$  0.00056 &   -37.950 &     0.023 &   1.329 $\pm$   0.032 &  107.0 &    5.2 \\
  0.00377 $\pm$  0.00038 &    -56.20 &      0.25 &    3.03 $\pm$    0.35 &    557 &    128 \\
  0.00375 $\pm$  0.00055 &    -65.76 &      0.16 &    1.35 $\pm$    0.23 &    111 &     38 \\
\hline\\[-1.5ex]
\multicolumn{6}{c}{B0834$-$196 (GMRT); ~~$N_{\rm comp} = ~7$; ~~$\chi_r^2=0.943$:}\\
\hline
  0.00124 $\pm$  0.00044 &     50.73 &      0.94 &     3.3 $\pm$     1.3 &    643 &    509 \\
   0.0120 $\pm$   0.0014 &     11.07 &      0.16 &    4.15 $\pm$    0.38 &   1045 &    186 \\
   0.0200 $\pm$   0.0015 &    11.177 &     0.042 &   1.197 $\pm$   0.092 &     87 &     13 \\
   0.0477 $\pm$   0.0034 &     2.739 &     0.022 &   0.784 $\pm$   0.046 &   37.3 &    4.2 \\
   0.0420 $\pm$   0.0058 &    -0.721 &     0.065 &    1.44 $\pm$    0.11 &    126 &     18 \\
   0.1414 $\pm$   0.0045 &     2.308 &     0.055 &   2.442 $\pm$   0.071 &    362 &     20 \\
   0.0081 $\pm$   0.0023 &      -2.8 &       1.8 &     3.8 $\pm$     1.3 &    896 &    583 \\
\hline\\[-1.5ex]
\multicolumn{6}{c}{B0906$+$430 (WSRT); ~~$N_{\rm comp} = ~3$; ~~$\chi_r^2=1.035$:}\\
\hline
  0.00262 $\pm$  0.00030 &     12.47 &      0.27 &    3.14 $\pm$    0.45 &    598 &    175 \\
  0.00355 $\pm$  0.00050 &      1.97 &      0.10 &    0.93 $\pm$    0.16 &     52 &     18 \\
  0.00113 $\pm$  0.00017 &       3.1 &       1.9 &    16.4 $\pm$     2.0 &  16320 &   4179 \\
\hline\\[-1.5ex]
\multicolumn{6}{c}{B1151$-$348 (GMRT); ~~$N_{\rm comp} = 10$; ~~$\chi_r^2=1.083$:}\\
\hline
  0.00217 $\pm$  0.00043 &    -35.66 &      0.50 &    2.15 $\pm$    0.72 &    279 &    194 \\
  0.00438 $\pm$  0.00044 &    -31.05 &      0.24 &    2.02 $\pm$    0.34 &    247 &     86 \\
  0.00442 $\pm$  0.00059 &    -18.62 &      0.17 &    1.22 $\pm$    0.24 &     90 &     37 \\
  0.00658 $\pm$  0.00052 &    -14.87 &      0.26 &    1.88 $\pm$    0.44 &    214 &    104 \\
  0.00872 $\pm$  0.00073 &    -11.28 &      0.19 &    1.76 $\pm$    0.31 &    187 &     68 \\
   0.0880 $\pm$   0.0027 &    -5.583 &     0.015 &   1.274 $\pm$   0.029 &   98.4 &    4.7 \\
   0.0381 $\pm$   0.0034 &    -4.198 &     0.079 &    4.27 $\pm$    0.17 &   1104 &     91 \\
   0.0430 $\pm$   0.0025 &    -2.425 &     0.019 &   1.021 $\pm$   0.047 &   63.2 &    6.1 \\
  0.02396 $\pm$  0.00057 &     6.605 &     0.029 &   1.303 $\pm$   0.043 &  102.9 &    7.1 \\
  0.00531 $\pm$  0.00053 &     10.08 &      0.13 &    1.46 $\pm$    0.20 &    129 &     37 \\
\hline
\end{tabular}\\
\end{center}
\label{table:fit2}
\end{table}
\setcounter{table}{0}
\begin{table}
\caption{({\it continued}) The parameters of the multi-Gaussian fits.}
\vskip -0.15in
\begin{center}
 \begin{tabular}{cr@{ $\pm$}rcr@{ $\pm$}r}
\hline
$\tau_{\rm peak}$ & \multicolumn{2}{c}{$V_c$} & $b$ & \multicolumn{2}{c}{$\Td$} \\
 & \multicolumn{2}{c}{km~s$^{-1}$} & km~s$^{-1}$ & \multicolumn{2}{c}{K}\\
\hline\\[-1.5ex]
\multicolumn{6}{c}{B1245$-$197 (GMRT); ~~$N_{\rm comp} = ~6$; ~~$\chi_r^2=1.006$:}\\
\hline
  0.00309 $\pm$  0.00094 &    -17.11 &      0.20 &    0.79 $\pm$    0.28 &     38 &     27 \\
  0.00296 $\pm$  0.00095 &    -12.19 &      0.22 &    0.82 $\pm$    0.32 &     40 &     32 \\
  0.00728 $\pm$  0.00067 &     -6.80 &      0.57 &    2.95 $\pm$    0.69 &    529 &    249 \\
   0.0243 $\pm$   0.0081 &     -3.39 &      0.11 &    1.45 $\pm$    0.20 &    127 &     36 \\
   0.0108 $\pm$   0.0034 &      -1.2 &       1.2 &    2.52 $\pm$    0.84 &    385 &    257 \\
  0.00373 $\pm$  0.00084 &      9.15 &      0.18 &    0.97 $\pm$    0.25 &     57 &     30 \\
\hline\\[-1.5ex]
\multicolumn{6}{c}{B1328$+$254 (WSRT); ~~$N_{\rm comp} = ~4$; ~~$\chi_r^2=1.027$:}\\
\hline
  0.00131 $\pm$  0.00016 &    -23.47 &      0.29 &    2.84 $\pm$    0.44 &    490 &    152 \\
  0.00132 $\pm$  0.00015 &     -3.29 &      0.28 &    3.04 $\pm$    0.40 &    559 &    149 \\
  0.00048 $\pm$  0.00015 &     11.23 &      0.74 &     2.8 $\pm$     1.0 &    483 &    361 \\
  0.00098 $\pm$  0.00023 &    -17.95 &      0.26 &    1.26 $\pm$    0.37 &     96 &     57 \\
\hline\\[-1.5ex]
\multicolumn{6}{c}{B1328$+$307 (WSRT); ~~$N_{\rm comp} = ~4$; ~~$\chi_r^2=1.025$:}\\
\hline
  0.00654 $\pm$  0.00022 &    -7.371 &     0.045 &   1.949 $\pm$   0.085 &    230 &     20 \\
  0.00730 $\pm$  0.00024 &   -14.281 &     0.029 &   1.122 $\pm$   0.047 &   76.3 &    6.5 \\
  0.00549 $\pm$  0.00021 &   -28.785 &     0.041 &   1.357 $\pm$   0.059 &    112 &     10 \\
  0.00109 $\pm$  0.00019 &     -8.41 &      0.68 &     9.2 $\pm$     1.0 &   5143 &   1168 \\
\hline\\[-1.5ex]
\multicolumn{6}{c}{B1345$+$125 (GMRT); ~~$N_{\rm comp} = ~5$; ~~$\chi_r^2=1.130$:}\\
\hline
  0.00298 $\pm$  0.00085 &    -48.06 &      0.14 &    0.59 $\pm$    0.19 &     21 &     15 \\
   0.0080 $\pm$   0.0010 &    -3.930 &     0.074 &    0.80 $\pm$    0.12 &     38 &     12 \\
   0.0799 $\pm$   0.0011 &    -1.173 &     0.011 &   1.454 $\pm$   0.020 &  128.2 &    3.7 \\
  0.00876 $\pm$  0.00099 &     -3.08 &      0.22 &    5.28 $\pm$    0.32 &   1691 &    220 \\
  0.00180 $\pm$  0.00053 &     19.52 &      0.37 &    1.52 $\pm$    0.52 &    140 &    101 \\
\hline\\[-1.5ex]
\multicolumn{6}{c}{B1611$+$343 (WSRT); ~~$N_{\rm comp} = ~2$; ~~$\chi_r^2=1.014$:}\\
\hline
  0.00177 $\pm$  0.00021 &     -4.10 &      0.40 &    4.14 $\pm$    0.56 &   1037 &    278 \\
  0.00093 $\pm$  0.00025 &    -30.32 &      0.56 &    2.54 $\pm$    0.80 &    390 &    244 \\
\hline\\[-1.5ex]
\multicolumn{6}{c}{B1641$+$399 (WSRT); ~~$N_{\rm comp} = ~1$; ~~$\chi_r^2=1.003$:}\\
\hline
  0.00082 $\pm$  0.00013 &     -1.25 &      0.61 &    4.72 $\pm$    0.86 &   1350 &    492 \\
\hline\\[-1.5ex]
\multicolumn{6}{c}{B1814$-$637 (ATCA); ~~$N_{\rm comp} = ~3$; ~~$\chi_r^2=1.027$:}\\
\hline
   0.3083 $\pm$   0.0051 &   -0.9024 &    0.0035 &  0.8638 $\pm$  0.0095 &   45.2 &    1.0 \\
   0.1088 $\pm$   0.0049 &    -1.055 &     0.017 &   2.052 $\pm$   0.051 &    255 &     13 \\
  0.00927 $\pm$  0.00092 &     -2.37 &      0.25 &    7.53 $\pm$    0.42 &   3433 &    388 \\
\hline\\[-1.5ex]
\multicolumn{6}{c}{B1827$-$360 (GMRT); ~~$N_{\rm comp} = ~9$; ~~$\chi_r^2=1.097$:}\\
\hline
  0.00152 $\pm$  0.00048 &    -15.29 &      0.32 &    1.27 $\pm$    0.49 &     97 &     78 \\
  0.00266 $\pm$  0.00080 &      -6.1 &       2.9 &      5.3 $\pm$    2.6 &   1678 &   1731 \\
   0.0057 $\pm$   0.0019 &     -3.86 &      0.27 &    0.98 $\pm$    0.28 &     59 &     35 \\
   0.0349 $\pm$   0.0017 &    -1.944 &     0.057 &   1.329 $\pm$   0.088 &    107 &     15 \\
   0.1371 $\pm$   0.0054 &     4.395 &     0.098 &    1.71 $\pm$    0.11 &    177 &     23 \\
   0.0640 $\pm$   0.0061 &      4.44 &      0.27 &    5.29 $\pm$    0.24 &   1699 &    165 \\
   0.0231 $\pm$   0.0076 &     4.017 &     0.042 &    0.73 $\pm$    0.11 &     32 &     11 \\
   0.1023 $\pm$   0.0086 &     6.865 &     0.090 &   1.596 $\pm$   0.078 &    154 &     16 \\
   0.0439 $\pm$   0.0013 &    10.961 &     0.019 &   1.076 $\pm$   0.032 &   70.2 &    4.3 \\
\hline
\end{tabular}\\
\end{center}
\label{table:fit3}
\end{table}
\setcounter{table}{0}
\begin{table}
\caption{({\it continued}) The parameters of the multi-Gaussian fits.}
\vskip -0.15in
\begin{center}
 \begin{tabular}{cr@{ $\pm$}rcr@{ $\pm$}r}
\hline
$\tau_{\rm peak}$ & \multicolumn{2}{c}{$V_c$} & $b$ & \multicolumn{2}{c}{$\Td$} \\
 & \multicolumn{2}{c}{km~s$^{-1}$} & km~s$^{-1}$ & \multicolumn{2}{c}{K}\\
\hline\\[-1.5ex]
\multicolumn{6}{c}{B1921$-$293 (GMRT); ~~$N_{\rm comp} = ~7$; ~~$\chi_r^2=0.995$:}\\
\hline
    0.191 $\pm$    0.015 &     4.736 &     0.016 &   1.012 $\pm$   0.030 &   62.1 &    3.7 \\
   0.0640 $\pm$   0.0083 &     3.163 &     0.030 &   0.736 $\pm$   0.048 &   32.8 &    4.2 \\
   0.0240 $\pm$   0.0028 &      4.43 &      0.26 &    6.16 $\pm$    0.39 &   2301 &    294 \\
    0.170 $\pm$    0.015 &     4.171 &     0.025 &   2.202 $\pm$   0.086 &    294 &     23 \\
  0.00800 $\pm$  0.00098 &     13.31 &      0.36 &    3.49 $\pm$    0.40 &    739 &    170 \\
  0.00279 $\pm$  0.00053 &    -13.96 &      0.30 &    1.89 $\pm$    0.42 &    218 &     96 \\
  0.00164 $\pm$  0.00020 &      42.4 &       1.3 &    13.2 $\pm$     1.9 &  10605 &   3046 \\
\hline\\[-1.5ex]
\multicolumn{6}{c}{B2050$+$364 (WSRT); ~~$N_{\rm comp} = 17$; ~~$\chi_r^2=0.902$:}\\
\hline
  0.00085 $\pm$  0.00025 &     -76.6 &       2.8 &    10.2 $\pm$     4.3 &   6246 &   5027 \\
  0.00085 $\pm$  0.00029 &     -39.7 &       2.2 &     7.6 $\pm$     3.5 &   3469 &   3041 \\
  0.00168 $\pm$  0.00047 &    -52.79 &      0.68 &    2.85 $\pm$    0.99 &    492 &    325 \\
  0.00224 $\pm$  0.00064 &     30.31 &      0.35 &    1.48 $\pm$    0.52 &    133 &     88 \\
  0.00689 $\pm$  0.00062 &    -64.59 &      0.13 &    1.85 $\pm$    0.21 &    208 &     44 \\
  0.00742 $\pm$  0.00054 &    -21.73 &      0.14 &    2.33 $\pm$    0.20 &    329 &     53 \\
  0.01135 $\pm$  0.00089 &   -12.214 &     0.088 &    1.53 $\pm$    0.16 &    142 &     28 \\
   0.0164 $\pm$   0.0014 &    -6.931 &     0.051 &   0.788 $\pm$   0.084 &   37.6 &    7.6 \\
   0.0973 $\pm$   0.0026 &    -1.600 &     0.023 &   1.637 $\pm$   0.044 &  162.5 &    8.4 \\
   0.0438 $\pm$   0.0036 &     1.625 &     0.048 &   1.299 $\pm$   0.097 &    102 &     15 \\
   0.1559 $\pm$   0.0043 &     8.202 &     0.012 &   0.901 $\pm$   0.025 &   49.2 &    2.6 \\
    0.188 $\pm$    0.034 &      9.14 &      0.25 &    3.55 $\pm$    0.30 &    764 &    122 \\
   0.0428 $\pm$   0.0067 &     14.17 &      0.14 &    1.16 $\pm$    0.17 &     81 &     23 \\
   0.0532 $\pm$   0.0064 &    15.823 &     0.089 &   0.961 $\pm$   0.080 &   56.0 &    8.8 \\
    0.032 $\pm$    0.022 &      14.5 &       3.9 &     5.3 $\pm$     2.0 &   1727 &   1254 \\
  0.00731 $\pm$  0.00069 &     25.24 &      0.17 &    1.97 $\pm$    0.26 &    236 &     58 \\
   0.0414 $\pm$   0.0031 &     -0.46 &      0.62 &    7.59 $\pm$    0.52 &   3488 &    451 \\
\hline\\[-1.5ex]
\multicolumn{6}{c}{B2200$+$364 (WSRT); ~~$N_{\rm comp} = 11$; ~~$\chi_r^2=0.952$:}\\
\hline
  0.00181 $\pm$  0.00056 &     -29.3 &       2.3 &     7.7 $\pm$     3.6 &   3632 &   3267 \\
   0.0051 $\pm$   0.0018 &    -17.68 &      0.24 &    0.83 $\pm$    0.35 &     42 &     34 \\
   0.0041 $\pm$   0.0011 &    -10.30 &      0.53 &    2.42 $\pm$    0.77 &    355 &    221 \\
   0.0182 $\pm$   0.0016 &   -20.592 &     0.080 &    1.22 $\pm$    0.13 &     90 &     19 \\
    0.192 $\pm$    0.020 &    -1.622 &     0.022 &   0.557 $\pm$   0.043 &   18.8 &    2.8 \\
    0.289 $\pm$    0.020 &    -0.918 &     0.094 &   1.449 $\pm$   0.084 &    127 &     14 \\
    0.218 $\pm$    0.012 &     1.041 &     0.097 &    4.18 $\pm$    0.12 &   1058 &     60 \\
    0.315 $\pm$    0.014 &     1.354 &     0.037 &   0.981 $\pm$   0.040 &   58.4 &    4.6 \\
   0.0757 $\pm$   0.0043 &     6.428 &     0.035 &   1.216 $\pm$   0.072 &     90 &     10 \\
   0.0228 $\pm$   0.0013 &    13.808 &     0.069 &   1.439 $\pm$   0.098 &    126 &     17 \\
   0.1031 $\pm$   0.0017 &    18.149 &     0.012 &   0.874 $\pm$   0.017 &   46.3 &    1.7 \\
\hline\\[-1.5ex]
\multicolumn{6}{c}{B2203$-$188 (WSRT); ~~$N_{\rm comp} = ~6$; ~~$\chi_r^2=0.996$:}\\
\hline
  0.06515 $\pm$  0.00062 &     7.802 &     0.010 &   1.053 $\pm$   0.018 &   67.3 &    2.3 \\
  0.02882 $\pm$  0.00054 &     4.948 &     0.027 &   1.437 $\pm$   0.043 &  125.2 &    7.5 \\
  0.00119 $\pm$  0.00034 &    -21.91 &      0.76 &     3.6 $\pm$     1.3 &    796 &    563 \\
  0.00140 $\pm$  0.00018 &      -3.6 &       2.1 &    15.5 $\pm$     2.8 &  14617 &   5183 \\
  0.00377 $\pm$  0.00063 &     10.23 &      0.15 &    0.93 $\pm$    0.23 &     52 &     25 \\
  0.00324 $\pm$  0.00096 &    -6.138 &     0.092 &    0.39 $\pm$    0.13 &    9.0 &    6.2 \\
\hline\\[-1.5ex]
\multicolumn{6}{c}{B2223$-052$ (GMRT); ~~$N_{\rm comp} = ~7$; ~~$\chi_r^2=1.034$:}\\
\hline
   0.0976 $\pm$   0.0025 &    -6.919 &     0.013 &   1.662 $\pm$   0.025 &  167.5 &    5.0 \\
   0.0419 $\pm$   0.0031 &     -4.41 &      0.13 &    4.81 $\pm$    0.18 &   1405 &    105 \\
   0.1039 $\pm$   0.0024 &    -3.943 &     0.010 &   1.236 $\pm$   0.019 &   92.6 &    2.9 \\
   0.0085 $\pm$   0.0012 &      2.34 &      0.14 &   1.330 $\pm$   0.212 &    107 &     35 \\
   0.0232 $\pm$   0.0024 &      5.93 &      0.24 &    2.23 $\pm$    0.16 &    302 &     44 \\
   0.0232 $\pm$   0.0036 &     7.238 &     0.046 &   1.059 $\pm$   0.097 &     68 &     13 \\
  0.00210 $\pm$  0.00044 &     11.27 &      0.40 &    1.57 $\pm$    0.57 &    149 &    110 \\
\hline
\end{tabular}\\
\end{center}
\label{table:fit4}
\end{table}

\label{lastpage}

\end{document}